\documentclass[twocolumn,showpacs,preprintnumbers,amsmath,amssymb,superscriptaddress]{revtex4}

\usepackage{graphicx}
\usepackage{dcolumn}
\usepackage{bm}
\usepackage{color}

\begin{document}
\newcommand{\norm}[1]{\ensuremath{| #1 |}}
\newcommand{\aver}[1]{\ensuremath{\langle #1 \rangle}}
\newcommand{\red}[1]{ {\color{red}#1 }}

\title{Temperature changes when adiabatically ramping up an optical lattice}

\author{Lode Pollet}
\email{pollet@itp.phys.ethz.ch}
\affiliation{Theoretische Physik, ETH Z{\"u}rich, 8093 Z{\"u}rich, Switzerland}

\author{Corinna Kollath}
\affiliation{Universit{\' e} de Gen{\` e}ve, 24 Quai Ernest-Ansermet, CH-1211 Gen{\` e}ve, Switzerland}
\affiliation{Centre de Physique Th\'eorique, Ecole Polytechnique, 91128 Palaiseau Cedex, France}
\author{Kris Van Houcke}
\affiliation{Vakgroep subatomaire en stralingsfysica, \\
  Proeftuinstraat 86,
   Universiteit Gent, Belgium}
 \affiliation{Department of Physics, University of Massachusetts, Amherst MA 01003}
\author{Matthias Troyer}
\affiliation{Theoretische Physik, ETH Z{\"u}rich, 8093 Z{\"u}rich, Switzerland}

\date{\today}

\begin{abstract}
When atoms are loaded into an optical lattice, the process of gradually turning on the lattice is almost adiabatic. In this paper we investigate how the temperature changes when going from the gapless superfluid phase to the gapped Mott phase along isentropic lines. To do so we calculate the entropy in the single band Bose-Hubbard model for various densities, interaction strengths and temperatures in one and two dimensions for homogeneous and trapped systems. Our theory is able to reproduce the experimentally observed visibilities and therefore strongly supports the view that current experiments remain in the quantum regime for all considered lattice depths with low temperatures and minimal heating.
\end{abstract}


\maketitle

\section{Introduction}
During recent years substantial progress has been made in cold atom experiments. Among the greatest achievements are the experimental realization of the superfluid to Mott insulating transition with bosonic atoms in optical lattices \cite{Greiner02}  and the BEC to BCS crossover in fermionic quantum gases \cite{Bloch_review07}. Due to the good tunability the excitement to use cold atomic gases as a so-called quantum simulator has been hard to temper. Such quantum simulators are systems which mimic -- in a controllable and ultraclean way --  simple strongly-interacting models such as the Hubbard model. With ultracold atom experiments the dream of using such quantum simulators to obtain new insight into the long-standing problems of other research areas seems `almost' feasible. One of the notorious challenges of condensed matter physics that might be solved in such quantum simulations is the fermionic Hubbard model \cite{Hofstetter,RVB}, believed to be relevant for high temperature superconductivity \cite{BednorzMuller}.
 
Beside the use as quantum simulators, the unprecedented tunability of cold atomic systems makes them very promising candidates for   quantum computers. The experimental production of entanglement has been a large step forward into performing quantum computations~\cite{Mandel03}.
 
The prerequisite of applying the quantum simulator to unknown physics, is a complete understanding of the present experiments. One would expect full agreement between the current generation of bosonic
experiments and theory, since the properties of the underlying homogeneous  Bose-Hubbard model are rather well studied. However, our understanding of the present experiments is far from complete. The interpretation of the results is mainly complicated by two points: 
\begin{itemize}

\item[(i)] The presence of an external trapping potential can cause the spatial coexistence of the superfluid and Mott phases. This replaces the quantum phase transition by a gradual crossover, which can obscure the characteristic signal of the quantum phases.

\item[(ii)] A change in temperature due to adiabatic or non-adiabatic origins can also hide the signature of the quantum phase transition, replacing it by a thermal transition.
%In particular, the application of an optical lattice
%potential could induce heating even in the ideal situation, i.e.~a totally
%isolated quantum gas and an adiabatic application of the potential. Ramping up the lattice is typically %slow ($\sim16$ ms per increase $E_{\rm R}$ in lattice laser intensity) compared to
%  the tunneling rate of the bosons ($\sim 1$ kHz, except for the deepest
%  lattices)~\cite{Gerbier05PRL}, and it is thus reasonable to assume that such
%  processes are adiabatic. 
Considerable heating in current experiments would cast serious doubt~\cite{Reischl05, Ho07, Diener07} on former interpretations. Knowledge of the temperature in an optical lattice is therefore highly desirable, but whereas the temperature of a weakly-interacting bosonic gas in a parabolic trap can be measured accurately, no reliable temperature measurement exists in the presence of a deep optical lattice. 

\end{itemize}

In this manuscript we reexamine the interpretation of recent experiments of bosons in optical lattices focusing on the effect of temperature in the presence of a trapping potential. We determine the lowest temperature that can be reached by loading the atoms adiabatically into the optical lattice. 
It is reasonable to assume that such processes are isolated and close to thermodynamic equilibrium at all times, since ramping up the lattice is typically slow ($\sim16$ ms per increase $E_{\rm R}$ in lattice laser intensity) compared to
  the tunneling rate of the bosons ($\sim 1$ kHz, except for the deepest
  lattices)~\cite{Gerbier05PRL}.
This assumption gives a lower bound for the temperature which can be achieved in the
experiments without applying further cooling techniques. 

We compare the results of our approach with the measured
experimental interference patterns over the whole parameter range. In contrast to
previous work \cite{Blakie03, Reischl05, Rey06, Ho07}, we can take the full range of
realistic heights of the  lattice potential and the full trapping potential
into account and our calculations are approximation-free and  based on first principles.

%Our main result is shown in Fig.~\ref{fig:adiab_visib}, where we have modeled
%experimental systems on a smaller two-dimensional lattice with a steep
%trapping potential. The visibility curve obtained along an adiabatic line and
%also taking the experimental procedure of calculating the visibility into
%account is in good agreement with experiment. We find
%that the visibility is not sensitive to temperature changes inside the quantum
%regime. Thus, adiabatic heating is of minor importance for the
%visibility. Temperature remains low for quadratic traps, a runaway temperature
%is not compatible with our numerical findings.
%\red{I would not like to reduce our results to this Fig! We should at least as
%well mention the density profiles along isentropic lines.}\\

After introducing the theoretical description of the quantum gas in section
\ref{sec:model}, we discuss the Quantum Monte Carlo methods we developed in
section \ref{sec:methods}. This section can be skipped by the
non-expert. In section  \ref{sec:onedim} we present results for the entropy in one dimension. We compare homogeneous systems with systems in the
presence of realistic trapping potentials across the superfluid to
Mott-insulating transition and we also study the Tonks-Giradeau gas. We make contact with
analytic approximations where possible, and we study the validity of the local
density approximation. Section ~\ref{sec:twodim} is dedicated to the entropy in
two-dimensional homogeneous and trapped systems. We study the influence of the
trapping potential, density, temperature, and of the actual experimental
procedure on the visibility in section ~\ref{sec:visibility}. We also compare the visibility obtained computationally with experimental data.

\section{Theoretical description}
\label{sec:model}
In order to prepare ultracold atoms into an optical lattice, the atoms are confined in an external trapping potential and cooled till they Bose-Einstein condense. Thereafter, additional lasers forming an optical lattice are turned on. One gradually increases the intensity of these lasers, slowly enough to remain in a low-energy state but also fast enough such that external influences are small.

Adiabaticity can be checked by reversing the process, i.e. decreasing the
intensity and comparing the state at a certain lattice height shows the same experimental
signature as the corresponding state while switching on the lattice. Such experiments indicate that the loading of the atoms might to a good approximation be considered as adiabatic.  

Atoms loaded in a sufficiently deep optical lattice are described by the Bose-Hubbard model~\cite{Jaksch98}, 
\begin{equation}
H = -t \sum_{\langle i,j \rangle} ^{L^d}{b}^{\dagger}_i{b}_j +
 \frac{U}{2}\sum_{i}^{L^d} {n}_i({n}_i-1) +  \sum_{i}^{L^d} \epsilon_i {n}_i, \label{eq:bose_hubbard_hamiltonian}
\end{equation}
The notation $\langle i,j \rangle$ refers to the sum over nearest-neighbor
sites only. Bosons are created on site $j$ by the operator $b_j^{\dagger}$ and the
number of bosons on site $j$ is counted by the number operator $n_j$. The
kinetic term describes hopping of the bosons with tunneling amplitude $t$,
while the on-site repulsion has strength $U$. We will work in the canonical
ensemble with a constant number of particles, $N$. The linear system size is $L$ and we work in dimensions $d=1,2$. We restrict ourselves to a
single-band Bose-Hubbard model, which is sufficient at low temperatures and lattice intensities around the superfluid-Mott transition, but it becomes approximative for very low repulsion or high
temperatures.
 
The external trapping potential is included using 
$\epsilon_i =  v_{\rm c}r^2$, where $v_{\rm c}$ describes the
strength of the trapping and $r$ the distance to the center of the lattice. 
If red detuned lasers are applied for creating the optical lattice potential the focusing of the lasers gives rise to an additional confinement. Under proper alignment of the two trapping potentials, the total trapping is given by $\epsilon(i) = m \omega^2 r^2/2$~\cite{GreinerPhD} with
\begin{equation}
\omega^2 = \omega_0^2 + 8V_0/(mw^2), 
\label{eq:laser_waist}
\end{equation} 
where $w$ is the waist of the lattice laser beam and $V_0$ its intensity
expressed in single-photon recoil energies, $E_{\rm R} =
h^2/2m\lambda^2$. Here $\lambda$ is the wavelength of the lattice laser beam
and $m$ is the mass of the atoms. We took the waists isotropic and neglected
corrections of the order of $\sqrt{V_0}$. The second term in Eq. (\ref{eq:laser_waist}) dominates already for moderately strong lattices.

The homogeneous Bose-Hubbard system shows a quantum phase transition from a superfluid to a Mott insulating state when the filling is commensurate~\cite{Fisher89}.  The transition occurs at $(U/t)_c = 3.28(4)$~\cite{PhD, Kuehner99} in one-dimensional systems, and  at $(U/t)_c = 16.74(1)$~\cite{Capogrosso07_bis, Elstner99} in two-dimensional systems. Whereas in the superfluid state  a continuous excitation spectrum exists at low energies, the Mott-insulator is characterized by a gap just above its ground state.

Analytically the entropy in the Bose-Hubbard model has been studied for
non-interacting \cite{Blakie03,Schmidt06} and weakly interacting
\cite{Blakie03,Rey06} bosons. The strongly interacting limit
for the Mott-insulator and the Tonks regime have been
considered for homogeneous \cite{Schmidt06,Rey06} and trapped systems \cite{Paredes04,
  Rey06}. 
In three dimensions, the entropy deep in the superfluid and Mott insulating phases was calculated using effective
masses~\cite{Capogrosso07}. Information on the trapped system was then
obtained using the local density approximation (LDA). The Bose-Hubbard model can only be solved analytically in these limiting cases,
where one is very deep in the superfluid or Mott insulating phase. Close to the phase
transition numerical tools have to be employed, such as the QMC simulations performed here.

%%%%%%%%Not really needed%%%%%%%%%%%
%We are interested in the
%exact results beyond LDA in trapped systems near the crossover from a purely
%superfluid system to a system with Mott shoulders and/or Mott domains by using
%quantum Monte Carlo methods. 

\section{Methods}
\label{sec:methods}

In this chapter we discuss a number of methods to determine the entropy 
\begin{equation}
S(\beta) = \beta (E - F) = \beta E + \ln(Z)
\end{equation}
where $E$ is the total (internal) energy of the system, $\beta=1/T$ the inverse temperature and $F = -\ln(Z)/\beta$ is the free energy. We use $k_B=1$. The main challenge is an accurate calculation of the partition function. It turns out that a combination of two methods discussed below gives the best results. Both are accurate, and do not involve any fitting nor noisy derivatives of numerical data, but they are efficient in a different temperature range. The results of both methods have been checked against each other for consistency. The flat histogram methods (Sec.~\ref{subsec:qwl})  works best at high temperatures, while the thermodynamic integration method (Sec.~\ref{subsec:tdintegration})  works better at low temperatures. This chapter is intended for the technically oriented readers and it is not necessary to read it in order to understand the discussion on the results in the next chapters.

\subsection{The canonical worm algorithm}\label{subsec:canworm}
 All our simulations employ a canonical worm algorithm. A worm algorithm~\cite{Prokofev98} in the path-integral representation is a quantum Monte Carlo algorithm where the decomposition of the partition function, 
\begin{eqnarray}\label{eq:zdecomp}
Z & = &  {\rm Tr} \sum_{n=0}^{\infty} \int_0^{\beta} dt_n \int_0^{t_n} dt_{n-1} 
\cdots
\int_0^{t_2} dt_1 \\ {} & {} & e^{-t_1 H_0}V
e^{-(t_2-t_1)H_0} 
\cdots e^{-(t_n - t_{n-1})H_0}V e^{-(\beta-t_n)H_0} \nonumber,
\end{eqnarray}
is sampled indirectly by making local moves in the Green function sector,  which is the extended configuration space of world lines with two open ends. Simulating the Bose-Hubbard model, we choose as diagonal
part $H_0$ the potential energy, while the hopping terms are the
perturbation $V$. 

In a canonical worm algorithm the operators of the equal-time Green function $b_i(\tau)b_j^{\dagger}(\tau)$ are propagated simultaneously. The extended partition function we sample reads
\begin{eqnarray}
Z_e & = & {\rm Tr} \left[ {\mathcal T} \left( \left(
  b_i(\tau)b_j^{\dagger}(\tau) + {\rm h.c.} \right) \exp(-\beta H) \right) \right] \nonumber \\
  {} & = & \sum_{n=0}^{\infty} \sum_{ \{ \vert  i_1 \rangle \}, \ldots, \{ \vert i_{n+1} \rangle \} }  \sum_{i,j} \int_0^{\beta} 
 \int_0^{t_n}  \cdots \int_0^{t_2} \cdots \nonumber \\
{} & {} &  W_{\rm e}(.) dt_1 \cdots d\tau \cdots dt_n \label{eq:weights},
\end{eqnarray}

where the terms $W_{\rm e}( . )$ denote
\begin{eqnarray}
W_{\rm e}(.) & = & e^{-t_1 E_1} \langle i_1|V|i_2\rangle e^{-(t_2-t_1)E_2} \cdots \nonumber \\
{} & {} & e^{- (\tau - t_{k}) E_k} \langle i_{k}| b_i^{\dagger}(\tau) b_j(\tau) |i_{k+1} \rangle e^{-(t_{k+1} - \tau) E_{k+1}}  \nonumber \\
{} & {} & \cdots e^{-(t_{n+1} - t_n)E_{n+1}}\langle i_{n+1} | V | i_1 \rangle e^{-(\beta-t_{n+1})E_1},
\end{eqnarray}
with $E_i=\langle i_1|H_0|i_1\rangle$, and we have introduced sums over a complete basis set between any two off-diagonal operators.
The terms $W_{\rm e} $ are all positive and can thus be used as weights in a Monte Carlo sampling.

An efficient updating scheme  has been presented in Ref.~\cite{Rombouts05,Vanhoucke05} and allows
the straightforward computation of  the kinetic and potential energy,
density, compressibility, equal time Green function, and superfluid density. However, the partition function is not a thermodynamic average and is harder to compute. 

\subsection{Flat histogram methods}\label{subsec:qwl}
The goal of a flat histogram method is to obtain a density of states $\rho(X)$ (where the coordinate $X$ in classical simulations is usually the energy)  directly by a random walk in $X$-space instead of performing a canonical simulation at fixed temperature. By sampling each value of $X$ with a probability ${\mathcal G}(X)\propto 1/\rho(X)$ we obtain a flat (constant) histogram $H(X) \rho(X) {\mathcal G}(X) = {\rm const.}$ 
%by performing Metropolis steps with acceptance ratio $\min \left[ 1, {\mathcal G}(X_j)/{ \mathcal G}(X_i) \right]$. $X_i$ and $X_j$ are the values of the coordinate before and after the Metropolis step and ${\mathcal G} (X)$ contains our current knowledge about the density of states. ${\mathcal G}(X)$ is modified during the simulation, which breaks detailed balance.  A histogram $H(X)$ is produced which tends to be flat in the long run, since $\rho(X) {\mathcal G}(X) = H(X)$ (if both sides are normalized). 

In the Wang-Landau sampling scheme \cite{Wang01},  a crude guess for ${\mathcal G}(X)$ is iteratively updated until it converges by a multiplicative factor $f$. During consecutive Wang-Landau iterations, $f$ is reduced according to $f \to \sqrt{f}$ when the current histogram $H(X)$ is considered to be sufficiently flat. Then the histogram is reset, we have a more accurate estimator for the density of states,  and the sampling restarts with the smaller $f$. The convergence of the scheme was proven by Zhou and Bhatt~\cite{Zhou03}. In particular, they showed that the minimum number of steps in each Wang-Landau iteration should scale as $1/\sqrt{f}$. The generalization of the Wang-Landau scheme to quantum systems was discussed in Ref.~\cite{Troyer03}. Here, we generalize Eq. (4) of Ref.~\cite{Troyer03} to the path-integral formulation:
\begin{eqnarray}
Z_{\lambda} & = & {\rm Tr} \exp \left( -\beta(H_0 - \lambda V \right) \nonumber \\ 
{} & = & {\rm Tr} \sum_{n=0}^{\infty} \int_0^{\beta} dt_n \int_0^{t_n} dt_{n-1}  \cdots \int_0^{t_2} dt_1 e^{-t_1 H_0}\lambda V \nonumber \\ 
{} & {} & e^{-(t_2-t_1)H_0} 
\cdots e^{-(t_n - t_{n-1})H_0}\lambda V e^{-(\beta-t_n)H_0} \nonumber \\
{} & = & \sum_{n = 0}^{\infty} g(n) \lambda^n.
\end{eqnarray}
The expansion order $n$ corresponds to the number of kinks (particle hoppings) present in the path
integral representation of a configuration. The original partition function
(which is a function of the inverse temperature $\beta$) can be found back by
setting $\lambda = 1$. The density of states $g(n)$ corresponds to
\begin{equation}
g(n) = \sum_{\vert  i_1 \rangle, \ldots, \vert i_n \rangle} \sum_{i,j} \int_0^{\beta} 
 \int_0^{t_n}  \cdots \int_0^{t_2} W(.)  dt_1 \cdots dt_n,
\end{equation}
where $W$ denotes the weight of a diagonal configuration, 
\begin{eqnarray}
W(.) & = &  e^{-t_1 E_1} \langle i_1 | V | i_2\rangle e^{-(t_2-t_1)E_2} \ldots \nonumber \\
{} & {} &  e^{-(t_n - t_{n-1})E_n} \langle i_n | V | i_1\rangle e^{-(\beta-t_n) E_1}.
\end{eqnarray} 
In such configurations all world-lines are continuous, and it occurs during the Monte Carlo run when the two open ends (worms) cancel each other on the same site and imaginary time. 

Using the canonical worm algorithm \cite{Rombouts05, Vanhoucke05}, the density of states can be obtained as follows : A single Monte Carlo step is defined from a diagonal to a new diagonal configuration and has an acceptance factor $q'$. Taking the density of states into account, the acceptance factor should be modified to
\begin{equation}
q( x \to y) = \min \left[ 1, g(x) q' / g(y) \right],
\end{equation}
where $g(x)$ is the density of states corresponding to the expansion order of the old configuration $x$. When the expansion order of the new configuration is larger than a predefined maximum expansion order, the update is rejected.

When the Wang-Landau iteration is finished, we can obtain the partition function for all values of $\lambda$. For the Bose-Hubbard model that means that we obtain a whole set of partition functions through the scaling $\beta t \to \beta t \lambda, U/t \to U/\lambda t$ (thus $\beta U$ is constant). A trap would also rescale as $V \to \lambda V$, which makes this scaling less useful in the trapped case and we use it just to obtain values at $\lambda=1$. If we had worked in the grand-canonical ensemble, the chemical potential would scale analogously, and we lose all control over the particle number. Here we see a distinct advantage of the canonical ensemble over the grand-canonical ensemble.

The normalization of $g(n)$ is fixed by calculating the partition function 
$Z_N$ in the canonical ensemble for the case 
without hopping ($t=0$). 

\subsection{Thermodynamic integration}\label{subsec:tdintegration}
The second method we discuss is the thermodynamic integration method. We choose a set of inverse temperatures
\begin{equation}
\beta_0 = 0 <  \beta_1 <  \beta_2 <  \ldots <  \beta_n.
\end{equation}
Then the partition function can be written as
\begin{equation}
\ln Z_{\beta_n} = \ln Z_{0} + \sum_{j = 1}^n \ln \frac{ Z_{\beta_j}}{Z_{\beta_{j-1}}} \label{eq:lnz_sum}
\end{equation}
The partition function at infinitely hot temperatures $Z_0 = Z_{\beta_0 = 0}$ can be found by solving the combinatorial problem of placing $N$ bosons on $L$ lattice sites,
\begin{equation}
\ln Z_0 = \ln \left( \begin{array}{c} L+N-1 \\ N \end{array} \right) = \sum_{j = L}^{L+N-1} \ln j - \sum_{j=1}^N \ln j.
\end{equation}
The ratios in Eq. (~\ref{eq:lnz_sum}) can be estimated through the weights introduced in Eq. (~\ref{eq:weights}), 

\begin{eqnarray}
\frac{Z_{\beta_{j-1}} } {Z_{\beta_j}} & = & \frac {\sum_{\sigma} W_{\beta_{j-1}} (\sigma)} {\sum_{\sigma} W_{\beta_j} (\sigma) } \nonumber \\
{} & = & \frac{\sum_{\sigma}   \frac{W_{\beta_{j-1}} (\sigma)} {W_{\beta_j} (\sigma)} W_{\beta_j} (\sigma)}   {\sum_{\sigma} W_{\beta_j} (\sigma) } \nonumber \\
{} & = & \left\langle  \frac {W_{\beta_{j-1}} (\sigma)} {W_{\beta_j} (\sigma) } \right\rangle_{\beta_j}.
\end{eqnarray}
where we sample all configurations $\sigma$ at the temperature $\beta_j$. In the canonical worm algorithm we have to measure
\begin{eqnarray}
\frac {W_{\beta_{j-1}} (\sigma)} {W_{\beta_j}} & = & \frac{\beta_{j-1}^n e^{-\beta_{j-1} E_d} }{ \beta_j^n e^{-\beta_j Ed}} \nonumber \\
{} & = & \left( 1 - \frac{\Delta \beta}{\beta_j} \right)^n e^{\Delta \beta E_d}, \label{eq:int_weightsratio}
\end{eqnarray}
with $\Delta \beta = \beta_j - \beta_{j-1}$ and $E_d$ the time averaged potential energy of the configuration. We can thus compute the partition function at $\beta_j$ if we know the partition function at $\beta_{j-1}$.  

The accuracy of the scheme depends on the overlap between the system at $\beta_j$ and the one at $\beta_{j-1}$. If the overlap is small, the error on the partition function will increase rapidly and  propagate systematically on to lower temperatures. In particular the first term $\beta_1$ should be chosen sufficiently close to zero, since there are only contributions if the expansion order is zero. 
  The fluctuations in the (diagonal) energy in Eq. ~(\ref{eq:int_weightsratio})  are exponentially hard to control. We therefore choose our set of values of $\beta$ such that $\Delta \beta E_d < 1$. At large values of $U$ or large particle numbers, more $\beta$-points are required. When we are close to the ground state things get easier since there energy fluctuations are suppressed. In the limit that $\Delta \beta$ is infinitely small, the scheme reduces to an energy integration. Note however that the scheme remains exact when $\Delta \beta$ is finite.

\subsection{Numerical strategy}
Although we tried a number of alternatives, none of them were satisfactory. We briefly make some remarks about them. We tried to use 
\begin{equation}
\frac{\partial S}{\partial U} = - \frac{\frac{1}{2} \partial \sum_i \langle n_i (n_i-1) \rangle} {\partial T}
\label{eq:double_occup}
\end{equation}
and integrate the density fluctuation with respect to inverse temperature, but the scatter of the data was much bigger than the trend-line, which made this approach prohibitive without an adequate fitting procedure.  Integrating the specific heat
\begin{equation}
S(\beta) = \int_0^{1/\beta} c_V(T')/T' dT',
\end{equation}\label{eq:int_spec_heat}
after differentiating the fitted curve through the energy was used in Ref.~\cite{Capogrosso07, Kato07, Werner05}. However, the division by the temperature is misbehaving at low $T$ and the specific heat computed via  the fluctuation formula $c_V = \beta^2 ( \langle E^2 \rangle - \langle E \rangle^2)$ is a quantity that converges slowly in the Monte Carlo simulation. We have also combined a grand-canonical directed loop algorithm in the 
path-integral representation \cite{Pollet07_LOWA}) with a quantum Wang-Landau reweighting scheme,
though the fact that the Wang-Landau reweighting also changes the
density, made this approach very cumbersome. A better attempt was developing a
canonical directed-loop algorithm in the stochastic series expansion \cite{SSE} and combining it with a
quantum Wang-Landau reweighting scheme. This approach has the advantage that
one obtains all temperatures down to the one corresponding to the pre-chosen cut-off length at once. The drawback is that in a SSE representation the large values of $U$ are also sampled, requiring  extremely large orders even for moderate temperatures.

We obtained the highest accuracy by combining the two methods
outlined in detail above. For high temperatures,  up to $\beta t \approx 0.5$, the
combination of the canonical worm algorithm with the flat histogram (QWL)
scheme is best and fast. For larger $\beta t$, and since we are interested in
the entropy for a very large number of $\beta t$ over a relatively small
temperature range, thermodynamic integration is best. This method
should only be used close to the ground state. Otherwise the fluctuations in
Eq. (\ref{eq:int_weightsratio}) are hardly controllable which might lead to
large systematic errors. \\

We compare the accuracy of both methods for a homogeneous one-dimensional system and for a trapped two-dimensional system in Table~\ref{table:comp_1d} and Table~\ref{table:comp_2d}, respectively. The data and the error bars shown in the tables do not reflect the full computational cost, since for the propagation method we need many more intermediate values of $\beta$ which are not shown in the tables. Similarly, for high $\beta$ the flat histogram method requires only a single Monte Carlo run, but the cost scales exponentially with $\beta$ (and the number of particles).\\

\begin{table}
\caption{Comparison between the flat histogram ('QWL') and direct integration ('chain') method for a homogeneous one-dimensional Bose-Hubbard system with $U/t=2$ and $N=45$.} 
\begin{tabular}{| l | l | l |}
\hline
$\beta t$ & $S_{\rm QWL}$ & $S_{\rm chain} $ \\
\hline 
\hline
$0.5$ & $44.2(2)$ & $44.08(3)$ \\
$1.0$ & $28.1(2)$ & $27.83(3)$ \\
$1.5$ & $19.0(3)$ & $18.6(1)$ \\
$2.0$ & $13.6(4)$ & $13.2(1)$ \\
$3.0$ & $7.9(4)$ & $ 7.6(6)$ \\
\hline
\hline
\end{tabular}
\label{table:comp_1d}
\end{table}

\begin{table}
\caption{Comparison between the flat histogram ('QWL') and direct integration ('chain') method for a trapped two-dimensional Bose-Hubbard system with $U/t=100$ and $v_c/t=2.5$.} \label{table:comp_2d}
\begin{tabular}{| l | l | l |}
\hline
$\beta t$ & $S_{\rm QWL}$ & $S_{\rm chain} $ \\
\hline
\hline 
$0.2$ & $ 36.9(3)$ & $36.7(2)$ \\
$0.5$ & $12.2(3)$ & $12.1(2)$ \\
$0.7$ & $7.2(4)$ & $7.4(4)$ \\
$1.5$ & $1.9(2)$ & $2.0(4)$ \\
\hline
\hline
\end{tabular}
\end{table}

\section{Isentropic lines in one dimensional systems}
\label{sec:onedim}

We will start our discussion of the one-dimensional case with a homogeneous lattice and compare our results with analytic results in limiting cases. We will then gradually make the discussion more realistic (and more complicated) by including the parabolic confinement. The first step is discussing the entropy in a system of constant quadratic trapping, $v_c/t = {\rm const.}$. The second step is
the case of an external parabolic trapping potential which is further
strengthened by the focus of the lattice laser beams as is currently done in
most experiments.  We will compute the entropy with these two confining
potentials starting in the superfluid and going to the Mott-insulator or the
Tonks-Giradeau gas. We will also check the quality of a numerically based local
density approximation and will find that in its regime of validity the speed-up in computing the entropy of trapped systems is considerable.

%************************************************************************************************
%
%************************************************************************************************
\subsection{Homogeneous system}

\begin{figure}
\centerline{\includegraphics[width=\columnwidth]{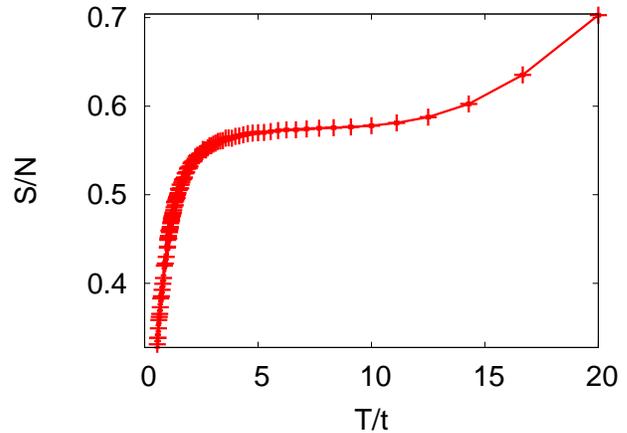}}
\caption{Entropy per particle as a function of temperature for a homogeneous one-dimensional Bose-Hubbard system with $U/t=100$, $N=40$ and $L=50$ ($n=0.8$). The plateau hints at the existence of a gap in the excitation spectrum.  }
\label{fig:plateau}
\end{figure}

The dependence of the entropy on the temperature is closely related to the energy spectrum. 
Weakly interacting superfluid systems have an continuous energy spectrum. Therefore the entropy rises continuously with temperature.
Mott insulating states have an energy gap just above the ground state and entropy is exponentially suppressed up to temperatures of the order of $k_B T/U \approx 0.1$~\cite{Rey06,Schmidt06}.
Above the gap the bands of excited states lead to a finite entropy if temperature is high enough.
Strongly interacting incommensurate systems also have gaps in the spectrum, but at substantially higher energies,  signaled by a plateau in the entropy in Fig.~\ref{fig:plateau}. \\

\begin{figure}
\centerline{\includegraphics[angle=270,width=\columnwidth]{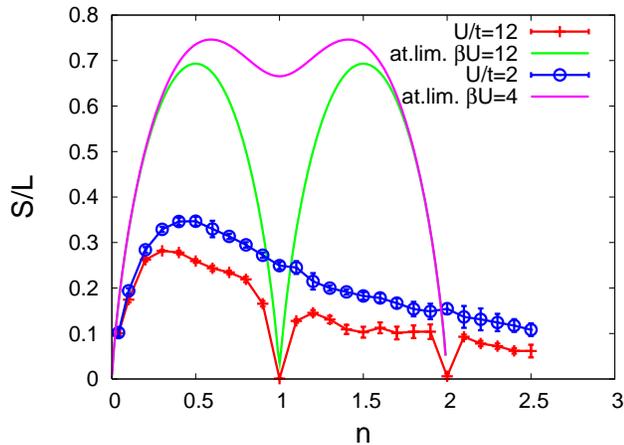}}
\caption{Entropy per site as a function of the filling factor for a homogeneous
  one-dimensional Bose-Hubbard system. QMC data are shown by the symbols for $U/t=12$ and $U/t=2$
  when $\beta t= 2$ and $L=50$. The lines are guide to the eyes. Comparison is
  made with the atomic approximation at different temperatures where only particle - hole (1p-1h)
  contributions are taken into account. }
\label{fig:S_1d_density}
\end{figure}

The dependence of the entropy per site on the filling is shown in
Fig.~\ref{fig:S_1d_density} for different interaction strengths at a
moderately low temperature $\beta t =2$. 
For weakly interacting systems ($U/t = 2$) the entropy reaches a maximum at half filling and decays monotonously for higher filling factors. Mott regions in strongly interacting systems ($U/t=12$) appear as strong dips in Fig.~\ref{fig:S_1d_density}, where the entropy is exponentially suppressed because the temperature is well below the gap.  We also make a comparison with the atomic limit approximation, where only single particle-hole excitations are taken into account.  The agreement with the Monte Carlo data is only qualitative, because the atomic limit misses higher order particle-hole excitations which are important for $U/t = 12$. The atomic limit is expected to work well near commensurability, but already for densities $0.9$ and $1.1$ the deviation with Monte Carlo is considerable. The atomic limit approximation also incorrectly predicts a symmetric curve around $n=1$. 
 
\begin{figure}
\centerline{\includegraphics[width=\columnwidth ]{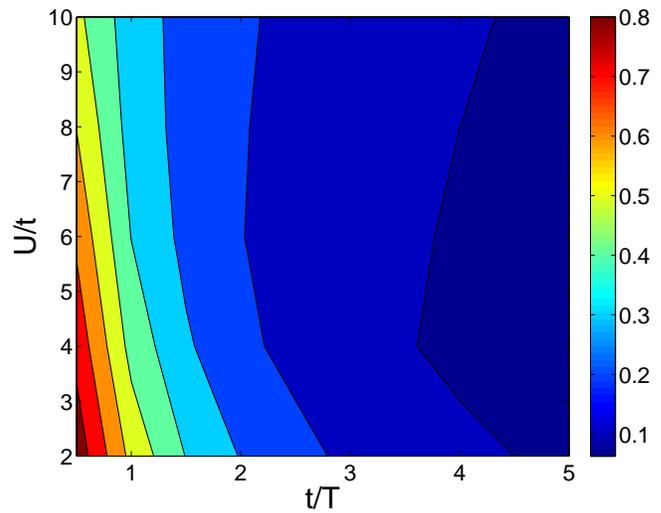}} 
\caption{Entropy per site  of a homogeneous one-dimensional Bose-Hubbard
  model. There are 40 particles and 50 sites at incommensurate filling. The
  system is always in the superfluid regime. The error on the entropy is less
  than two percent in all cases. This representation of the entropy allows to read off the final
  temperature easily when adiabatically ramping up the lattice. For instance,
  if the initial temperature is $\beta t=2$ for $U/t=2$ then the final
  temperature is $\beta t=1.3$ for $U/t=10$. } \label{fig:S40}
\end{figure}

Figure \ref{fig:S40} shows the entropy per site for a system of 40 particles on a
lattice of 50 sites as a function of the interaction strength $U/t$ and inverse
temperature $t/T = \beta t$. Since we are away from integer filling  we are always in the superfluid phase. 
At intermediate and high temperatures $\beta t< 3$, the entropy decreases when
increasing the interaction strength. In contrast at lower temperature, $\beta t\gtrsim 3$, we see a non-monotonic behavior of the entropy with a
minimum close to the critical interaction strength in a commensurate system. 
The same qualitative behavior was observed for a smaller lattice of $L=20$ sites. The reason is the presence of the nearby quantum phase transition to a commensurate Mott state. The mass of the quasi-hole decreases when going away from the tip of the Mott lobe at $(U/t)_c$ to higher values of $U/t$. The quasi-hole absorbs more entropy and the entropy thus increases.
Moving along
isentropic lines we observe in units of $t$ some mild heating when the initial temperature is
reasonably high, $\beta t< 3$, but cooling if the initial temperature is sufficiently low, $\beta t > 4$.

\begin{figure}
\centerline{\includegraphics[width=\columnwidth ]{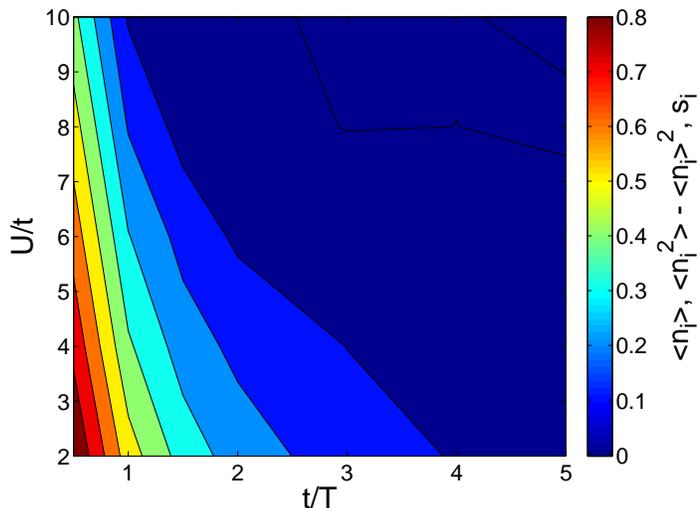}} 
\caption{Entropy per site   of a homogeneous one-dimensional Bose-Hubbard
  model at commensurate filling. There are 50 particles and 50 sites. For $U/t
  < 3.28$~\cite{PhD, Kuehner99} the system is in the superfluid phase in the thermodynamic limit at $T = 0$, for larger $U/t$ a gapped Mott state is formed. The error on the entropy is of the order of $0.05$, making it impossible to observe the exponential decay of the entropy in the Mott state.} \label{fig:S50}
\end{figure}

The commensurate case in Fig.~\ref{fig:S50} shows the same behavior as
Fig.~\ref{fig:S40} for low values of $U/t<(U/t)_c$ when we remain in the superfluid regime. However, when the Mott state develops, i.e.~$U/t>(U/t)_c$, the gap in
the spectrum opens and the entropy at constant temperature decreases considerably. Along adiabatic trajectories the temperature in units of $t$ shoots up near the transition point. This has been discussed for the homogeneous system in
Ref.~\cite{Schmidt06}, from which the authors deduced  that the temperature in present
experiments must be of the order of $U$. However, Rey et al. \cite{Rey06} pointed out that in the presence of a parabolic confining potential less heating occurs in hard-core bosonic systems than in a homogeneous system. The next section addresses the same question for soft-core bosons.

%************************************************************************************************
%
%************************************************************************************************

\subsection{Entropy distribution in a constant parabolic trapping potential}

\begin{figure}

\centerline{\includegraphics[width=\columnwidth ]{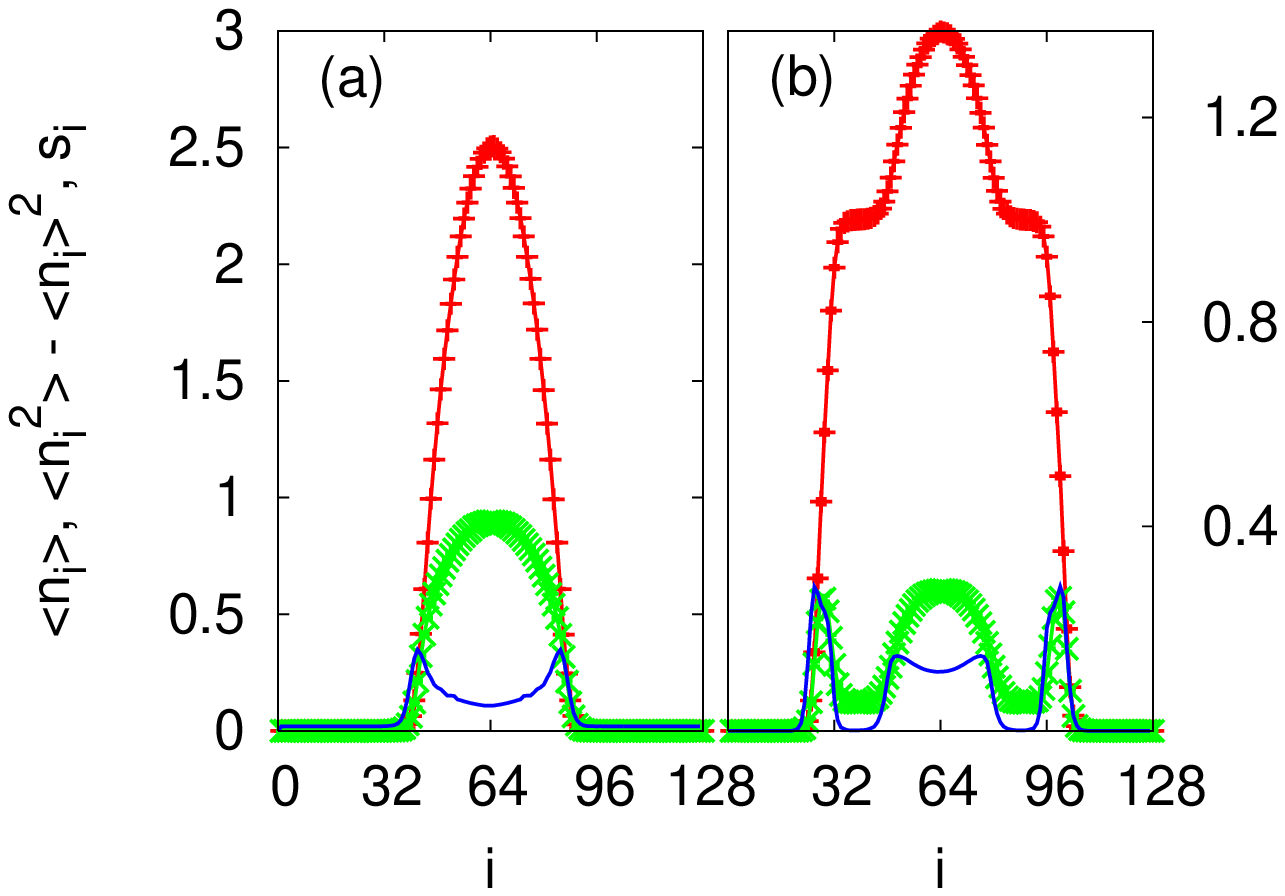}} 

\caption{ Density profile (red line, top curve in the center), density variance (green line, middle curve in the center)
  and entropy profile  (blue line, lowest curve in the center) for a one-dimensional trapped system for different parameters.  a) $U/t = 2$, inverse temperature ${\beta t=2}$, system size $L = 128$, the
  trapping potential $v_c/t =
  0.00829$, and the density $ N = 80 $. The total entropy calculated by the
  LDA approximation is $S_{\rm LDA}=10.75$ and by QMC simulations $S_{\rm QMC}=9.8(1)$.
b) $U/t = 12$, inverse temperature ${\beta t=2}$, system size $L = 128$, the
  trapping potential $ v_c/t =
  0.00829$, and $N=80$ particles. $S_{\rm QMC} = 9.2(1)$ from QMC and $S_{\rm LDA}=8.3$
  from LDA.  }
\label{fig:LDA}
\end{figure}

In the presence of an external trapping potential spatially separated quantum phases can coexist. 
In Fig.~\ref{fig:LDA} we show two density profiles in the presence of a
trapping potential. The first is for a superfluid state (Fig.~\ref{fig:LDA}a)) in which the density closely follows the form of the trapping potential. The large local variance
$\kappa=\aver{n_i^2}-\aver{n_i}^2$  demonstrates the number fluctuations in the superfluid state. 
The second density profile in Fig.~\ref{fig:LDA}b) is for a state in which the central inhomogeneous region
is surrounded by a Mott-insulating shell with commensurate filling and an
outer incommensurate region. The variance shows a clear suppression of the density
fluctuations in the Mott-insulating regions. 

The coexistence of spatially separated quantum phases can be understood in terms of a 
site-dependent chemical potential, the so-called local density approximation
(LDA). Physical quantities are determined by using on each site the results
obtained for a homogeneous system with the corresponding chemical potential.  The site-dependent effective chemical potential provides a scan through the phase diagram. This
approximation has been shown to work nicely for such quantities as the density or
the variance in regions where the trapping
potential varies slowly \cite{Wessel04,Bergkvist04}. However, the LDA breaks down
for steep trapping potentials and near the edges of Mott plateaus where numerical simulations are necessary to obtain reliable values \cite{Wessel04}. 
To get a better understanding of the entropy distribution in an inhomogeneous
system we developed a canonical and improved variant of the LDA, dubbed iLDA:
 we first calculate the exact density profile using a full QMC simulation for
 the trapped simulation. Then we take for
 every site the entropy from a homogeneous run corresponding
 to that density. This variant has the advantage that we start from the
 exact density profile, taking into account the rounding near the edges of the Mott plateaus due to the finite gradient of the trap. We tested the approach by comparing the total entropy of the trapped system calculated as the
 sum of the single sites to full numerical simulations. We found, as shown in Tab.\ref{table:compare_LDA}, that the iLDA
 can capture the qualitative trend of full calculations, but cannot reproduce the
 exact values.

\begin{table}
\caption{Entropy comparison between iLDA and exact Monte Carlo results for different one-dimensional system parameters. The trapping potential for the parameters in the third row is $v_c/t = 0.00829$.} \label{table:compare_LDA}
\begin{tabular}{| l | r | r |}
\hline
parameters & $S$ by QMC & $S$ by iLDA \\
\hline 
Fig.~\ref{fig:LDA}, left panel & $9.8 \pm 0.1$ & 10.8\\
Fig.~\ref{fig:LDA}, right panel & $9.2 \pm 0.1$  & 8.3\\
 $U/t=6$, $\beta t=1$,$N = 80$ &
$22.7 \pm 0.1$ & 21.8\\
Fig.~\ref{fig:trap1d_dens} , $U/t=12$ & $ 7.9 \pm 0.2 $ & $6.6$\\
\hline
\end{tabular}
\end{table}

Using iLDA we obtained the entropy profiles for the density profiles of
Fig.~\ref{fig:LDA}. 
%If the whole system is superfluid (Fig.~\ref{fig:LDA} (a)) the entropy varies
%only slowly from site to site and shows maxima at the filling close to
%$n\approx 1/2$. 
In a system where superfluid and Mott-insulating regions
coexist, we clearly see that the
Mott-like regions are not able to accommodate entropy and the whole entropy
is in the superfluid regions. 
If the whole system is superfluid (Fig.~\ref{fig:LDA} (a)) the entropy varies
only slowly from site to site and shows maxima at the filling close to $n\approx 1/2$.
Since the filling is larger than one in the center a dip in the entropy
profile develops. 

\begin{figure}
\centerline{\includegraphics[width=\columnwidth ]{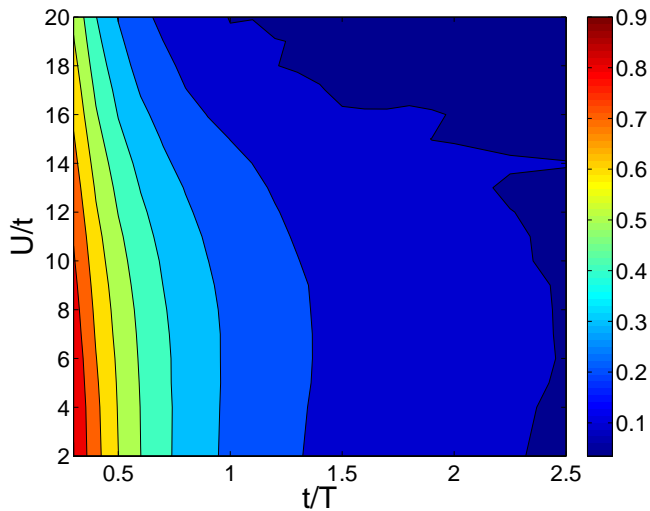}} 
\caption{Entropy  per particle  of a one-dimensional Bose-Hubbard model with
  constant trapping $v_c/t = 0.00829$. There are 80 particles and 128
  sites. The magnitude of the errors is approximately $0.1 - 0.2$. The
  roughness of the lowest isentropic line is within the error bars. } \label{fig:trap1d}
\end{figure}

\begin{figure}
\centerline{\includegraphics[angle=270,width=\columnwidth ]{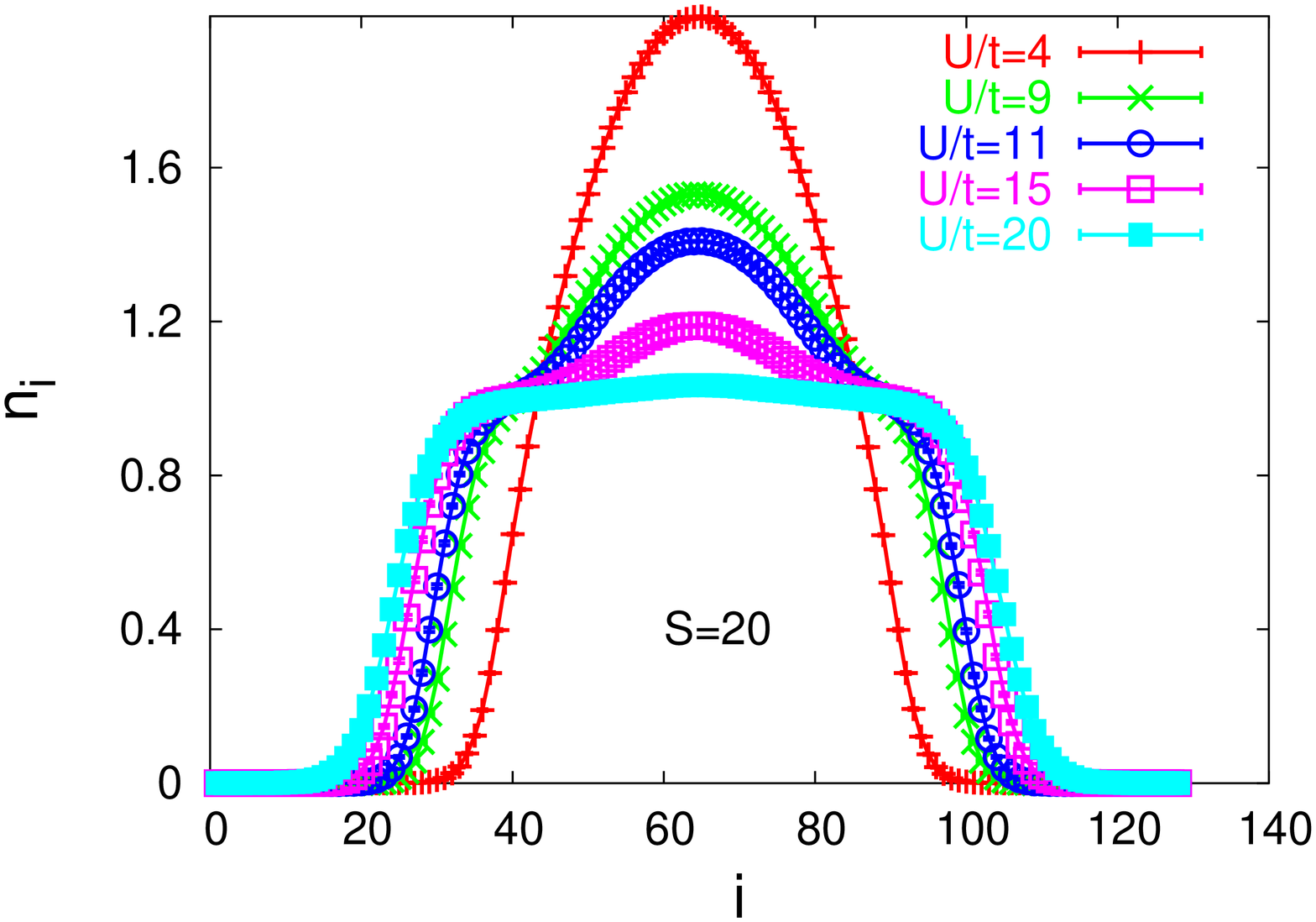}} 
\centerline{\includegraphics[angle=270,width=\columnwidth ]{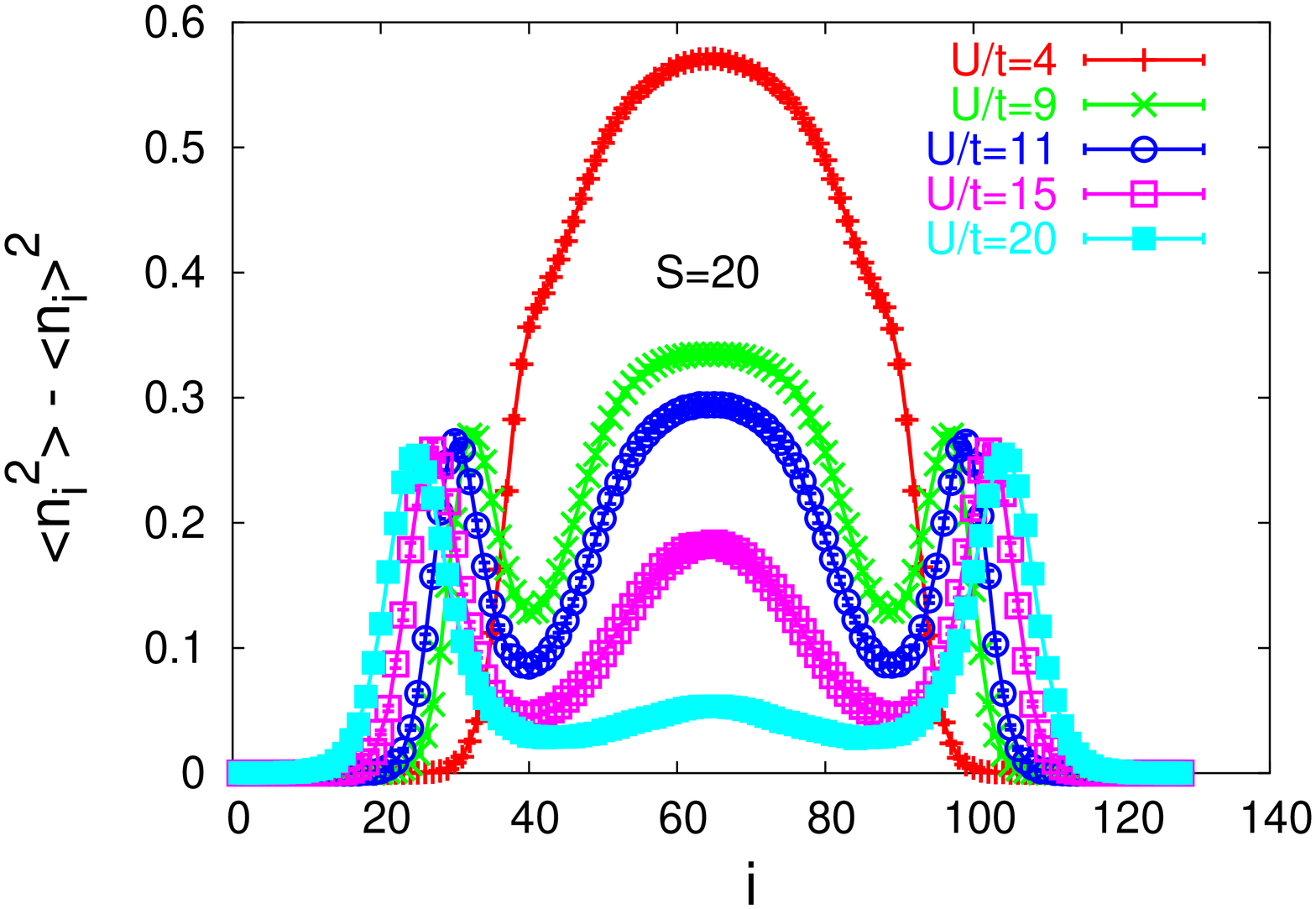}} 
\caption{(a) Density profiles and (b) variance along the isentropic line of $S=20$ in
  Fig.~\ref{fig:trap1d}. See Table~\ref{table:S_1d_consttrap} for the corresponding temperatures. 
} 
\label{fig:trap1d_dens}
\end{figure}

\begin{table}
\caption{Values of $U/t$, $\beta t$ and $U \beta$ along an adiabatic line  $S=25$ for the same parameters as in Fig.~\ref{fig:trap1d}. See Fig.~\ref{fig:trap1d_dens} for the corresponding density profiles. }
\label{table:S_1d_consttrap}
\begin{tabular}{| l | r |r |}
\hline
$U/t$ & $ \beta t$  & $\beta U$\\
\hline 
4  & 1.10(2) & 4.4(1) \\
9  & 1.10(2) & 9.9(2) \\
11& 1.00(5) & 11.0(5) \\
13 & 0.93(3) & 12.1(4) \\
15 & 0.80(5) & 12.0(8) \\
18 & 0.65(3) & 11.7 (5) \\
\hline
\end{tabular}
\end{table}

In Fig.~\ref{fig:trap1d} we show the entropy calculated in a full QMC calculation for a
constant trapping potential. In Table~\ref{table:S_1d_consttrap} we show the values of temperatures following an
isentropic line as extracted from the data. The concept of 'adiabatic heating' is complicated by the different energy scales (and units) used in the literature.  It is important to define with respect to which energy scale the temperature is measured as illustrated in Table~\ref{table:S_1d_consttrap}.  
We see that along isentropic lines in the
superfluid phase $(U /t < 10)$ temperature in units of the hopping $t$ remains roughly constant. At higher
temperatures there is some small heating in units of $t$, while at low temperatures we observe a little cooling. 

This behavior can qualitatively be understood by looking at the density profiles and its variance along isentropic lines for different values of $U/t$, as shown in Fig.~\ref{fig:trap1d_dens}.  Prior to the
formation of a wide commensurate region, the temperature remains
almost constant, for instance  $\beta t = 1.1$ for $U/t = 4$ and $U/t=9$ and $\beta t=1.0$ for $U/t = 11$. Only when a considerable volume percentage of the system turns insulating the incommensurate edges cannot accommodate the entropy anymore which results in a rise of the
  temperature, i.e.~$\beta t= 0.8$ for $U/t = 15$ to $\beta t= 0.7$ for $U/t = 20$.
 We checked that this effect is seen for a wide range of initial temperatures.

In units of the interaction strength $U$ a cooling takes
place. In particular, we see that for the chosen initial temperature the final temperature with respect to $U$ stays below the temperature for which  excitations in the
Mott insulator are created in a homogeneous system. Therefore, the Mott insulator is
stable up to the considered lattice height and the superfluid regions take most of the excitations.

%************************************************************************************************
%
%************************************************************************************************

\subsection{Entropy distribution in a realistic parabolic trapping potential}

\begin{figure}
\centerline{\includegraphics[width=\columnwidth ]{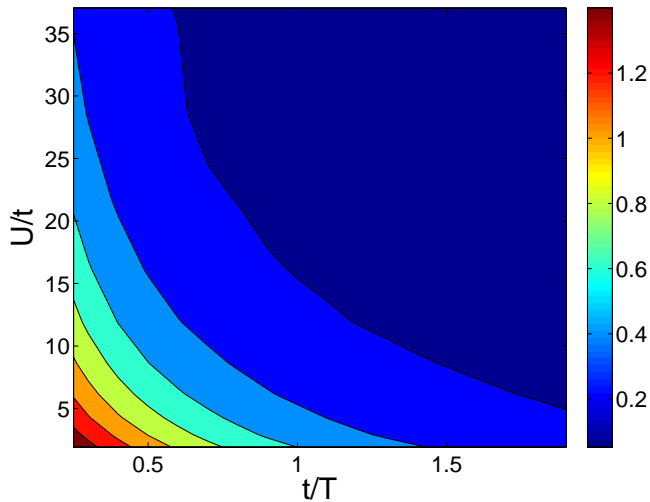}} 
\caption{Entropy per particle  of a one-dimensional Bose-Hubbard model with varying trapping potential, taking the waist of the laser into account according to Eq. (\ref{eq:laser_waist}) with a magnetic trapping frequency $\omega_0 = 2 \pi \times 30$ Hz and a laser waist $w = 160 \mu{\rm m}$ for ${}^{87}{\rm Rb}$ atoms. Data points were calculated for $V_0 = 1, 2, \ldots , 10 E_{\rm R}$ or for $ U/t = 1.9, 2.9, 4.3, 6.1, 8.6, 11.9, 16.1,  21.7, 28.6,$ and $ 37.0$ and for $\Delta \beta t \sim 0.1 $ or $0.2$.  There are 60 particles and 128 sites. The magnitude of the errors is a few percent.  } 
\label{fig:trap1dwaist}
\end{figure}

\begin{table}
\caption{Values of $U/t$, $\beta t$ and $\beta U$ along an isentropic line $S=13$ for the same parameters as in Fig.~\ref{fig:trap1dwaist}. }
\label{table:S13_N60}
\begin{tabular}{| c | r | r |r |r |r |r |}
\hline
$V_0[E_{\rm R}]$ &	$U/t$& $v_c/t$ & 	$\beta t$ &	$\beta U$ \\
\hline
4&	6.13&	0.008	& 1.62(4)&	9.9(2) \\
5&	8.62&	0.011	& 1.35(4) &	11.6(4) \\
6&	11.90&	0.015	& 1.10(4)&	13.1(4) \\
8&	21.74&	0.028	& 0.70(4)&	15.2(8) \\
9&	28.57&	0.037	& 0.61(2)&	17.5(6) \\
10&	37.04&	0.049	& 0.49(2)&	18.2(7) \\
\hline
\end{tabular}
\end{table}

Figure \ref{fig:trap1dwaist} shows the entropy when the finite waist of the optical lattice beam is taken into
account (and $v_c/t$ is no longer a constant but a function of the lattice
  laser intensity). Table~\ref{table:S13_N60}  gives the parameters for the temperature extracted along the isentropic line $S=13$.
We see that along this line the temperature in units of $t$ increases while in
units of $U$ the temperature decreases. This means that the formation of the
Mott-insulator is still possible starting at a low enough temperature, since the incommensurate regions take a lot of entropy.

\begin{figure}
\centerline{\includegraphics[angle=270,width=\columnwidth ]{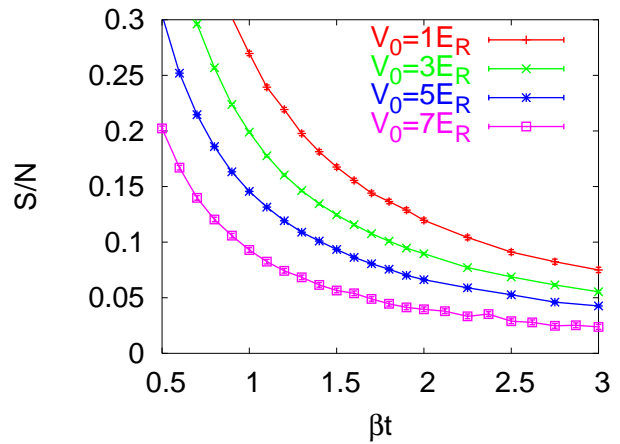}} 
\caption{Entropy per particle as a function of the inverse temperature $\beta$ for
  different intensities $V_0$ (expressed in recoil energies) of the lattice
  laser.  The system has $N=140$
  particles on a lattice of $L=128$ sites. We took a laser waist of $w = 160 \mu {\rm m}$. An isentropic line of $S=10$ goes
  through $\beta=3$ for $V_0=1$ and through $\beta=1$ for $V_0= 8
  $.  See Table~\ref{table:N140_S20} for temperatures along an adiabatic line $S=20$.  } 
\label{fig: entropy_trap_highocc}
\end{figure}

\begin{figure}
\centerline{\includegraphics[angle=270,width=\columnwidth ]{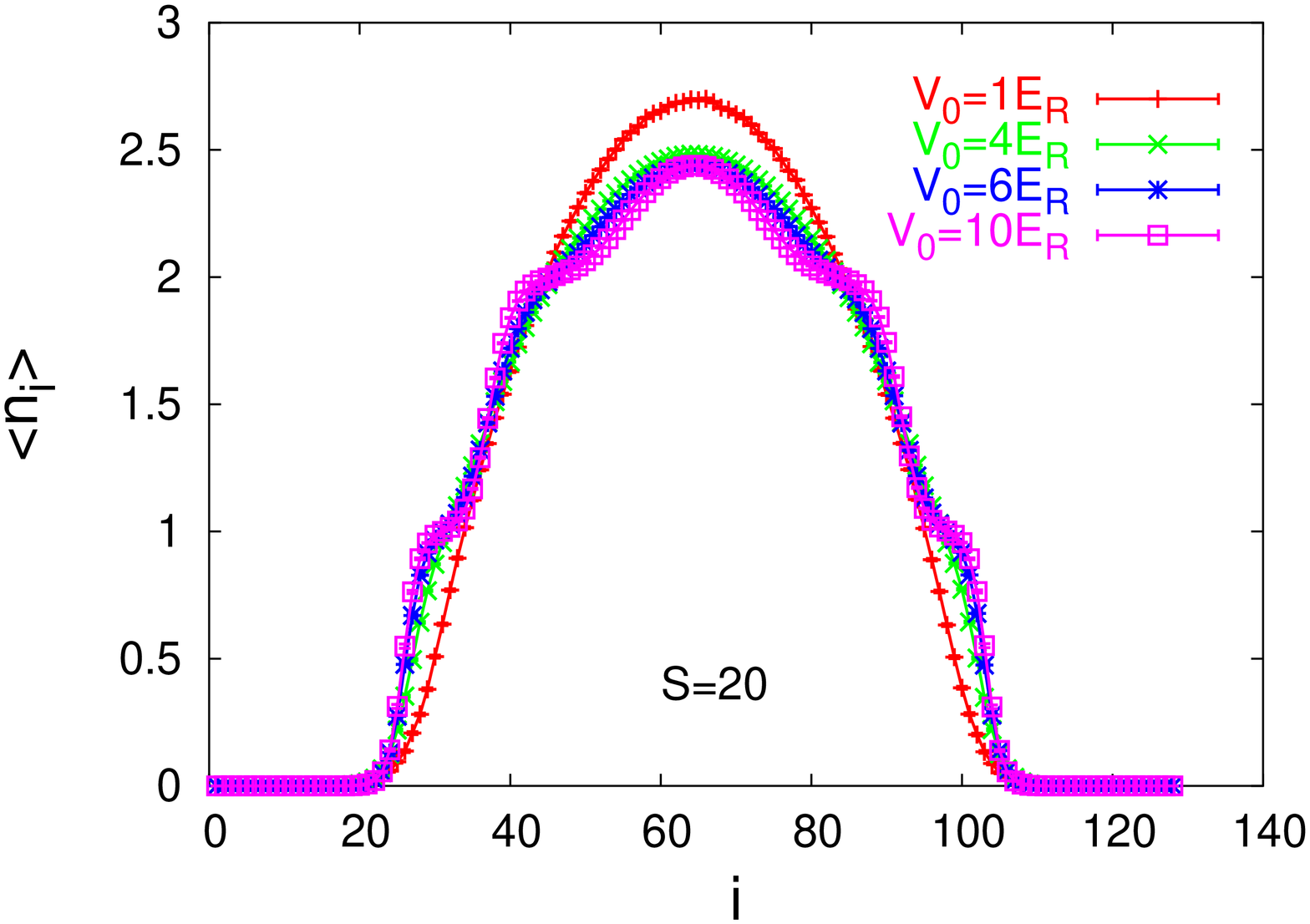}} \centerline{\includegraphics[angle=270,width=\columnwidth ]{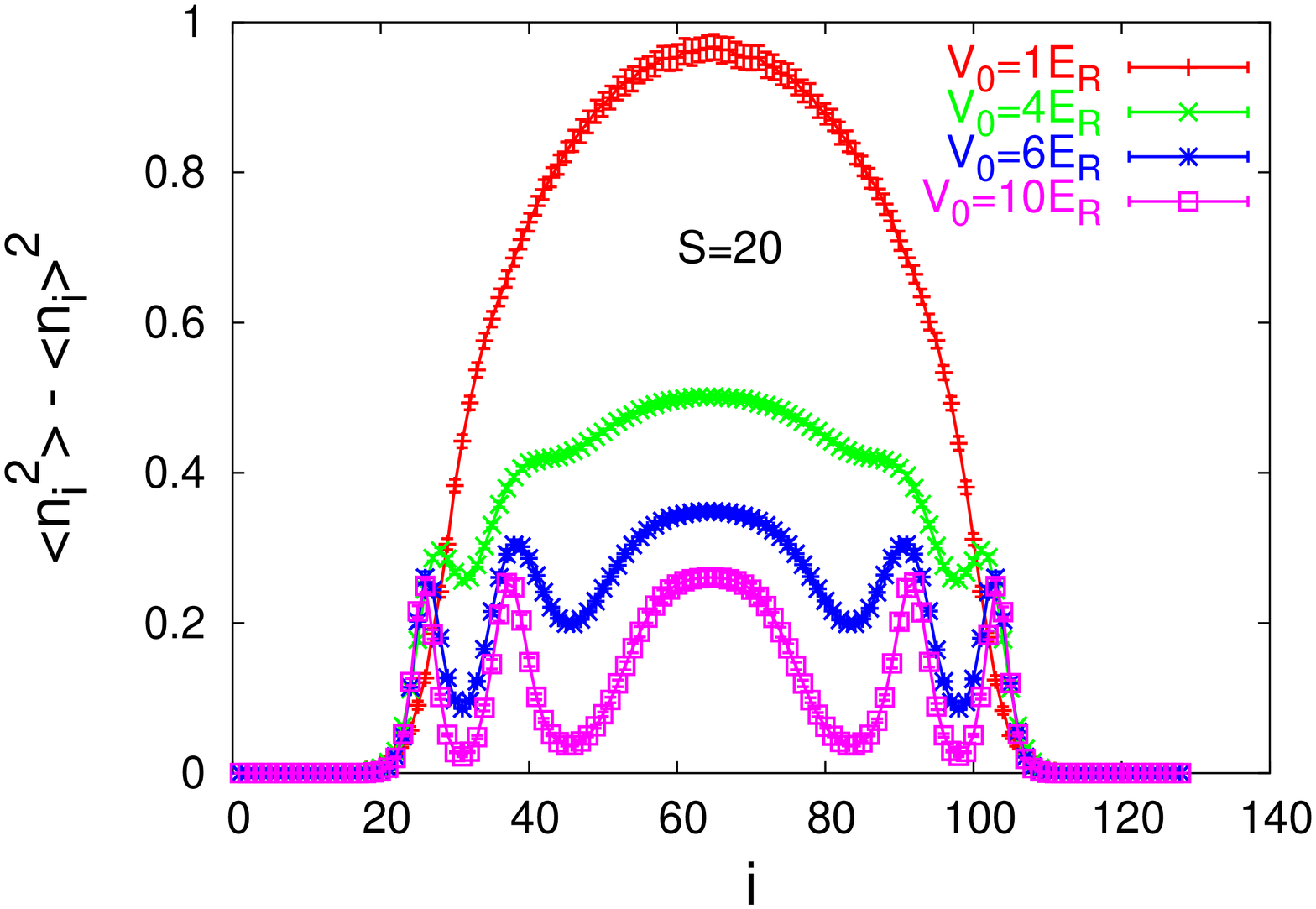}}
\caption{(a) Density profiles and (b) density variance profiles along the isentropic line of $S=20$ in
  Fig.~\ref{fig: entropy_trap_highocc}.  See Table~\ref{table:N140_S20} for the corresponding temperatures.
  %% The temperature rises from $\beta t = 1.7$ at $V_0 = 1 $ ${\rm E}_{\rm R}$ 
%%   over $\beta t= 1.1$ at $V_0 = 4 $ E$_{\rm R} $ to $\beta t = 0.8$ at $V_0 = 6 $ E$_{\rm R}$ and $\beta t = 0.26$ at $V_0 = 10 {\rm E}_{\rm R}$. 
  } 
\label{fig:trap1d_dens_waist_N140}
\end{figure}
\begin{table}
\caption{Values for $U/t$, $\beta t$, and $\beta U$ along an isentropic line
  $S \approx 20$ for the parameters as in Fig.~\ref{fig: entropy_trap_highocc}.  See Fig.~\ref{fig:trap1d_dens_waist_N140} for the corresponding density profiles.}
\label{table:N140_S20}
\begin{tabular}{| c | r |r |r |}
\hline
$V_0[E_{\rm R}]$ & $U/t$ & $t \beta$  & $U\beta$\\
\hline 
1 &1.92 & 1.70(2) & 3.2(1) \\
4 & 6.13 & 1.15(5) & 7.0(3) \\
5 & 8.62  & 1.00(3) & 8.6(3) \\
6 &11.90  & 0.85(5) & 10.2(6) \\
7 & 16.13 & 0.70(2) & 11.3(3) \\
8 & 21.74 & 0.55(4) & 12.0(9) \\
9 &  28.57 & 0.46(2) & 13.1(6) \\
% 10 : calculation is wrong since big neg entropy found for lower temperatures, old number cannot be trusted.
\hline
\end{tabular}
\end{table}

Figure \ref{fig: entropy_trap_highocc} shows the entropy as a function of
the temperature for different strengths of the optical lattice
potential. Compared to Fig.~\ref{fig:trap1dwaist}  the number of atoms is
increased to $N=140$. This has the consequences that even at low temperature
only small Mott-insulating regions can form and an incommensurate region
survives in the center of the trap (Fig.~\ref{fig:trap1d_dens_waist_N140}). In
Table~\ref{table:N140_S20} we show the temperature along the isentropic line $S= 20$. In units of $t$ we now get very strong heating, even in the
superfluid phase, due to the high occupation which does not accommodate as much entropy as the low occupation region (cf. Fig~\ref{fig:S_1d_density}). 
However, in units of $U$ we again find a temperature
decrease. 

One sees that the Mott-insulator is stable against the
temperature change, since the temperature stays below $0.1U$. Since the Mott-insulating regions are small and a broad
incommensurate region survives in the center, the Mott transition
does not play a central role in the behavior of the entropy curves.

\begin{table} 
\caption{Values for $U/t$, $\beta t$, and $\beta U$ along an isentropic line $S \approx 50$ for the parameters as in Fig.~\ref{fig: entropy_trap_highocc}. }
\label{table:N140_S50}
\begin{tabular}{| c  | r |r |r |r |r |}
\hline
$V_0[E_R]$ &	$U/t$ &	$\beta t$ &	$\beta U$ \\
\hline
1	& 1.916 	& 0.80(2) & 1.5(1) \\
3	& 4.23 	& 0.60(2) & 2.5(1) \\
4	& 6.14	& 0.50(2) & 3.1(1) \\
%5	& 8.62	& 0.50(2) & 4.3(2)  \\
7	& 16.13	& 0.27(2) & 4.4(5) \\
8	& 21.74	&0.22(2) & 4.8(4) \\
%10	& 37.037	& 0.11(1) & 4.0(4) \\
\hline
\end{tabular}
\end{table}

Starting at a higher initial temperature $\beta t= 0.8$ and following the
isentropic line $S\approx 50$, the same qualitative effect of temperature
increase in the units of $t$ and decrease in the units of $U$ can be
seen in Tab. ~\ref{table:N140_S50}. However, at $U/t \approx 20$, the temperature is still so high ($k_BT>0.1U$) that
no clean Mott-insulating region can be formed. 

%************************************************************************************************

%************************************************************************************************
%
%************************************************************************************************

\subsection{Tonks gas : one-dimensional trapped case}

\begin{figure}
\centerline{\includegraphics[width=\columnwidth ]{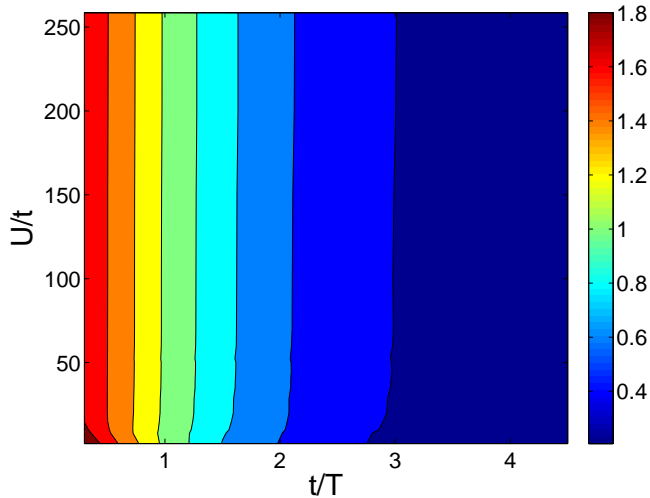}} 
\caption{Entropy per particle  of a one-dimensional Bose-Hubbard model with constant trapping $v_c/t = 0.00829$, approaching the Tonks regime for large values of $U/t$. There are $N = 15$ particles in the system of size $L=50$; the system is so dilute such that no Mott region is formed. } \label{fig:tonks}
\end{figure}

\begin{figure}
\centerline{\includegraphics[width=\columnwidth ]{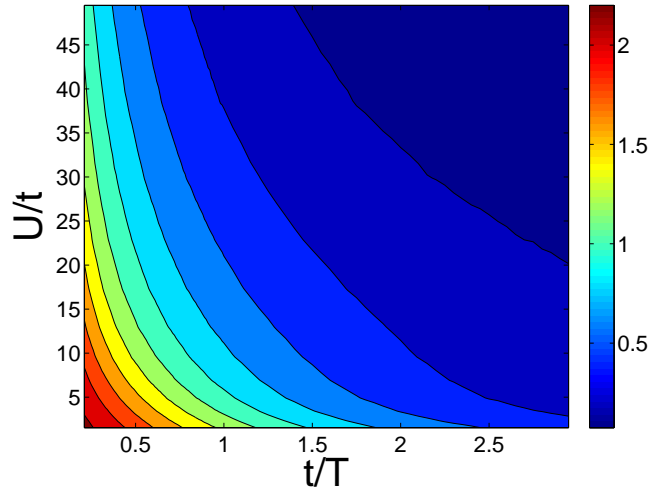}} 
\caption{Entropy per particle  of a one-dimensional Bose-Hubbard model with
  changed trapping, approaching the Tonks regime for large values of $U/t$. There are $N = 15$ particles in the system of size $L=50$} \label{fig:tonks_waist}
\end{figure}

When the potential energy between the atoms of a one-dimensional Bose-Hubbard model increases, the particles behave more and more like hard-core bosons. The limit of infinite repulsion and no multiple occupancies of a site, is called the Tonks-Giradeau gas. Quantities such as the energy, average density, variance of the density can be computed accurately by assuming non-interacting fermions. Other quantities such as the density matrix map to a non-interacting fermionic density matrix up to a phase factor coming from the Jordan-Wigner transformation. The experimental detection of the Tonks gas has demonstrated one of the fundamental concepts of quantum mechanics, namely the absence of a clear meaning of statistics in one-dimensional systems.

 The Tonks regime has been observed with~\cite{Paredes04} and without  a lattice~\cite{Weiss05}. 
In the experiment with a lattice, the data was analyzed using fermionization, and good agreement with experiment was found in the region $U/t>5$. The fermionization results were obtained at different temperatures  along adiabatic lines. Temperature rose from $\beta t = 2$ for a lattice depth of $V_0 = 4.6 E_{\rm R}$ till $\beta t = 0.77$ for $V_0 = 12 E_{\rm R}$ as a consequence of the change in $v_c/t$ when ramping up the lattice. Consistently, it was argued in Ref.~\cite{Reischl05}  that the temperature in the Tonks gas in a trap was of the order of the hopping $t$. 

The Tonks problem was also studied by Monte Carlo simulations~\cite{Pollet04, Wessel05_Tonks}  at a low but constant temperature $\beta t =1$, which is of the same order as the one used in the fermionization approach. The authors compared hard-core and soft-core bosons for a homogeneous lattice and for constant trapping. They found a gradual cross-over and found that the presence of a trap did not qualitatively change the Tonks onset. 

The contrast in the experimental interference pattern was almost completely gone for the deepest lattices~\cite{Paredes04}. Although this is consistent with a strongly repulsive superfluid gas, one might fear that similar patterns are produced by either a Mott state or a thermal state due to a combination of soft-core bosons and an increased trapping depth.

Our analysis is again carried out in two steps.
First, we calculated the entropy of a one-dimensional superfluid in the very
low density limit trapped in a constant parabolic trap. From
Fig.~\ref{fig:tonks} we deduce that temperature remains remarkably
constant. Thus, for a superfluid in the low density regime, adiabatic
processes are (almost) isothermal when the external trapping is constant and
weak. 
%This is reminiscent of superfluid Helium-4 where the isobaric
%compressibility is zero \red{citation is missing}.
Second, we make the simulation of the experiment more realistic. We
numerically evaluate the Bose-Hubbard parameters using the tight-binding
approximation, and obtain the same parameters $U$ and $t$ as in
Ref.~\cite{Pollet04}. In contrast to Ref.~\cite{Pollet04}, we now also
calculate the trapping parameter $v_c/t$ from the total axial trapping
$\omega_{\rm ax} = 2 \pi \times 60$ Hz ~\cite{Paredes04} for all optical lattice
depths. 
We find in Fig.~\ref{fig:tonks_waist} some heating, even in the low-density superfluid phase. Along an adiabatic line similar to the one taken in Ref.~\cite{Paredes04}, we find that the temperature increases from $\beta t= 2$ at $V_0 = 5 E_{\rm R}$ ($U/t = 6.9$) to $\beta t = 0.69$ at $V_0 = 12 E_{\rm R}$ ($U/t = 49.5$). Thus, the temperature increase compares very well to the one calculated assuming hard-core bosons~\cite{Paredes04}.

Summarizing, we confirm that the experiment has indeed reached the gradual cross-over toward the Tonks regime. The temperature remains of the order of the hopping $t$, even though a temperature increase of a factor of 3 is found due to the change in confinement strength $v_c/t$ when ramping up the lattice. It is exactly this increase in temperature that prevents the Mott domains from developing, since for $V_0 = 12 E_{\rm R}$ ($U/t = 49.5$, $v_c/t = 0.073$ ) a broad Mott domain appears in the center of the trap for 15 particles and $\beta t = 2$. At a temperature $\beta t = 0.69$ the central density is $0.9$ however.

%************************************************************************************************
%
%************************************************************************************************

\section{Results in two dimensions}
\label{sec:twodim}

\subsection{Homogeneous two-dimensional superfluid}

\begin{figure}
\centerline{\includegraphics[angle=270,width=\columnwidth]{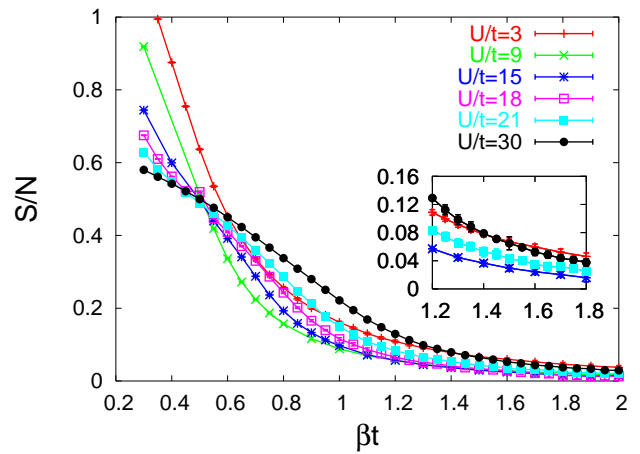}}
\caption{Entropy per particle for a system of size $L=20 \times 20$ in the superfluid phase, $N=320$. For a commensurate system, the transition happens at $(U/t)_c = 16.74(1)$~\cite{Capogrosso07_bis, Elstner99}. For low temperatures, we see an initial heating with increasing $U/t$, but around the transition point the presence of the Mott phase is felt and the system starts to cool, thanks to the lower effective mass. For larger values of $U/t$ we see a further cooling for low temperatures since the Mott phase is far away and we go deeper inside the superfluid phase. The inset shows the entropy for low temperatures (same axes and symbols as in the main figure). }
\label{fig:entropy_hom_2d}
\end{figure}

\begin{figure}
\centerline{\includegraphics[angle=270,width=\columnwidth]{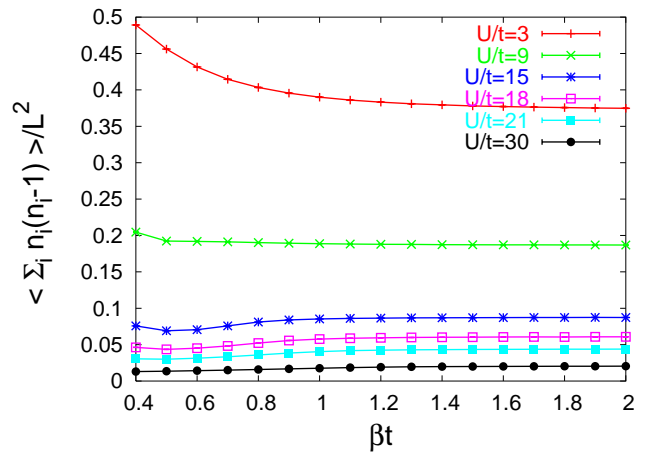}}
\caption{Double occupancies for the same system as in Fig.~\ref{fig:entropy_hom_2d}. The curves bend down for low temperatures at low values of $U/t$. For large values of $U/t$ a minimum is reached around $\beta t \approx 0.5$ while for larger values of $\beta t$ the curves bend slightly upwards with temperature. }
\label{fig:double_occup_hom_2d}
\end{figure}

We again start our analysis in two dimensions with the homogeneous case. The phase
transition from the superfluid to the Mott phase occurs for a $n=1$
commensurate system at $(U/t)_c = 16.74(1)$~\cite{Capogrosso07_bis, Elstner99}. We calculate the
entropy for a superfluid system close to commensurability, $n=0.8$. In
Fig.~\ref{fig:entropy_hom_2d} the dependence of the entropy on the temperature
and the interaction strength is shown. Its main behavior is similar to the one in a one-dimensional homogeneous
case as reported in Fig.~\ref{fig:plateau}.  
At fixed interaction strength the entropy shows only a small increase for low
temperatures. However, at a certain temperature it starts to increase strongly
before it bends down again and a plateau is formed. This can as in the 1D case
be related to the underlying band-structure in which first excitations in the
lowest band can be created. Above a certain temperature the corresponding gap
in the energy band-structure causes an intermediate saturation before at even
higher temperatures further bands can be excited.  

 As a function of the interaction strength, the entropy shows -- for constant but low temperature -- a minimum close to the superfluid to
Mott-insulator transition point of a commensurate system (inset of Fig.~\ref{fig:entropy_hom_2d}). As in the one-dimensional case this can be attributed to the effective mass change of the quasi-hole which has its maximum close to the phase transition point for the homogeneous system.

 The density fluctuation shown in Fig.~\ref{fig:double_occup_hom_2d} is consistent with the behavior of the entropy shown in Fig.~\ref{fig:entropy_hom_2d}, using the relationship of Eq. (\ref{eq:double_occup}). For infinitely hot temperatures the density fluctuation is independent of $U/t$ (not shown). For low values of $U/t$ density fluctuation goes down monotonously with $\beta t$. The normal-superfluid transition happens around $\beta t \approx 0.30(5)$ in our system of density 0.8 for a small lattice of $20 \times 20$ and $U/t = 3$. This transition belongs to the Kosterlitz-Thouless university class, and was studied in detail for the commensurate case in Ref.~\cite{Capogrosso07_bis, Elstner99}. 
 
 For large values of $U/t$ we enter the quantum (thermal) critical regime determined by the quantum critical point $U/t = 16.74$ ~\cite{Capogrosso07_bis, Elstner99} and a minimum in the density fluctuations is reached. This is clearly seen for $U/t = 15$ around $\beta t = 0.5$. For larger values of $U/t$ the minimum is reached at lower values of $\beta$. After this minimum is reached, the density fluctuations go slightly up with $\beta t$. 
% We are here in a normal state, the superfluid phase is not reached for temperatures shown  in the figure ({\color{red} In our small system we found the normal-superfluid transition around $\beta t \approx XXX$ for $U/t=30$}). 
The increase in the density fluctuations can be understood from the tendency of a dilute gas of vacancies (with respect to the $n=1$ Mott state as a vacuum) trying to condense. 
 %This is analogous to what happens with a weakly-interacting dilute Bose gas, where the vacuum state is the state without particles : The chemical potential shift is reduced from $2nU$ to $nU$ when going from the normal to the condensed phase. 
In our canonical simulation, we will observe a tendency to increase the number of vacancies which will enhance pair formation. Thus theoretically the density fluctuations contain a lot of information about the system, but from the practical point of view the almost flat slopes in the quantum regime make this quantity a bad candidate for thermometry.

%************************************************************************************************
%
%************************************************************************************************

\subsection{The superfluid to Mott insulating transition in a parabolic trap}

\begin{table}
\caption{Bose-Hubbard parameters chosen in the two-dimensional trapped system.} \label{table:pet_params}
\begin{tabular}{| l | r |}
\hline
$U/t$ & $v_c/t$ \\
\hline 
$5$ & 0.2 \\
$10$ & $0.33$\\
$15$ & $0.47$ \\
$20$ & $0.61$ \\
$25$ & $0.74$ \\
$50$ & $1.11$ \\
$100$ & $2.50$ \\
\hline
\end{tabular}
\end{table}

In two dimensions we only consider trapping potentials taking into account the influence of the finite waist of the lattice laser on our sample. We use the parameters shown in Table.~\ref{table:pet_params}. There are $N=200$ bosons and the total system size is $L=20 \times 20$.  These parameters are a compromise between increasing the trapping frequency while still obtaining meaningful results on a lattice of size $L=20 \times 20$. The filling in the system with $N=200$ atoms for weak interactions is chosen such that it is close to $n=2$ in the center for large $U/t$. 

\begin{figure}
\centerline{\includegraphics[angle=270,width=\columnwidth ]{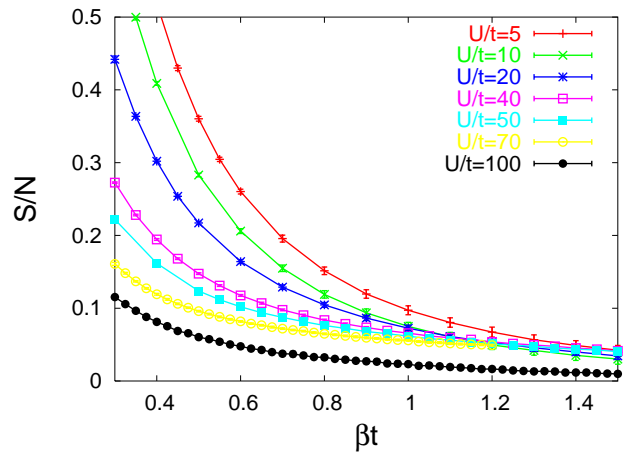}} 
\caption{Entropy per particle as a function of inverse temperature for a set of different
  values of $U/t$ with a trapping potential according to
  Table~\ref{table:pet_params}.  The curves for the intermediate curves
  coalesce within error bars for $\beta t > 1.5 $. The curve for $U/t=100$ is
  significantly below the other curves (see however text) up to temperatures
  of the order $\beta t = 2$. 
   } \label{fig:entropy_pet}
\end{figure}

The entropy dependence on temperature is shown in
 Fig.~\ref{fig:entropy_pet} for different values of $U/t$. For
 low values of the temperature the slope for the curves is very flat. For
 higher temperature a clear increase in the entropy in the system can be
 seen. For low temperature the curves for different $U/t$ almost coincide and
 only the curve for very strong interactions, i.e.~$U/t=100$, shows a
 considerably lower value of the entropy.

\begin{figure}
\centerline{\includegraphics[angle=270,width=\columnwidth ]{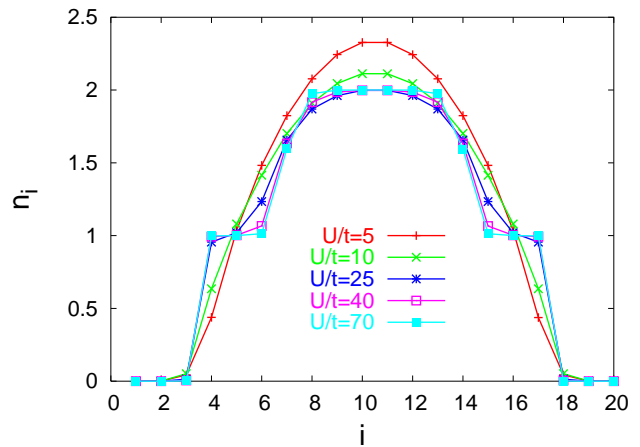}} 
\caption{One-dimensional cut of the density profiles for $y=L/2$ depending on
  the $x$ coordinate (labeled with 'i')  along an adiabatic line $S=10$. Error bars are smaller than the point size. See Table~\ref{table:pet_params} for the corresponding values of the trapping potential and Table~\ref{table:T_isentrop} for the corresponding temperatures.}
\label{fig:adiab_dens_s10}
\end{figure}

\begin{figure}
\centerline{\includegraphics[angle=270,width=\columnwidth ]{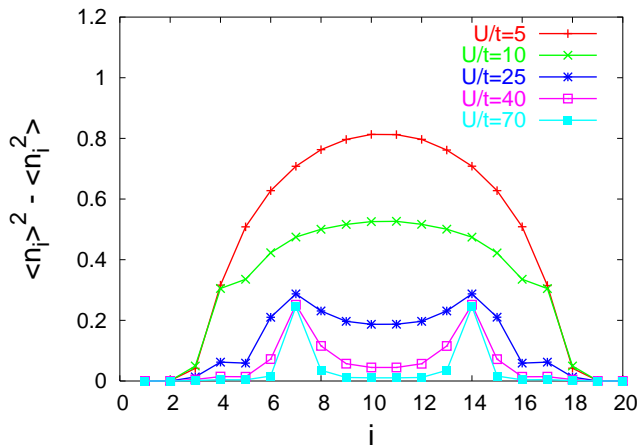}} 
\caption{Variance of the density along the $x$ direction (labeled with 'i')  for $y=L/2$ along an adiabatic line $S=10$, corresponding to the density profiles shown in Fig.~\ref{fig:adiab_dens_s10}. Error bars are smaller than the point-size. See Table~\ref{table:pet_params} for the corresponding values of the trapping potential and Table~\ref{table:T_isentrop} for the corresponding temperatures.}
\label{fig:adiab_compr_s10}
\end{figure}

The behavior of the entropy can be explained by considering the density
profiles of the system, shown in Fig.~\ref{fig:adiab_dens_s10} along an adiabatic line $S=10$. By increasing the optical lattice
potential the central density goes down and we see the appearance of the $n=1$ Mott-insulating region
at $U/t\approx 25$. These regions can be identified in the
variance profile in Fig.~\ref{fig:adiab_compr_s10}.
For stronger interactions, $U/t \sim 50$,  small
Mott-insulating regions with $n=1$ and $n=2$ exist. For $U/t=70$ the Mott-insulating
regions cover already a large volume fraction of the system as clearly signaled in
the variance profile. The formation of the large Mott-insulating region causes
the value of the entropy for the curve at $U/t=100$ to lie below the others
at low temperature. 

\begin{table}
\caption{Temperature along three adiabatic lines for the two-dimensional trapped Bose-Hubbard model with parameters according to Table~\ref{table:pet_params}. }
\label{table:T_isentrop}
\begin{tabular}{| l | r | r | r | r | r | r | }
\hline
$U/t$ & $\beta_{S=10}t$ & $\beta_{S=10}U$ & $\beta_{S=40}t $ &$\beta_{S=40}U $
& $\beta_{S=300}t$ & $\beta_{S=300}U$ \\
\hline 
5   & 1.40(5) & 7 & 0.70(2)& 3.5 & 0.13(1)& 0.65 \\
10 & 1.20(5) & 12 & 0.60(2) & 6 & 0.08(1)& 0.8 \\
25 & 1.30(5) & 32 & 0.50(2) & 12.5 &0.06(1) & 1.5 \\
40 & 1.25(5) & 50 & 0.40(2) & 16 & 0.06(1) & 2.4 \\
50 &  1.20(5) & 60 & 0.35(5) & 17.5&0.06(1)& 3 \\
70 &  1.18(2)   & 83& 0.25(1) & 17.5& 0.03(1)&2.1 \\
100 & 0.58(3) &58 & 0.18(2) & 18& 0.01(1)& 1 \\
\hline
\end{tabular}
\end{table}

The temperature evolution along isentropic lines for different initial temperatures is shown more quantitatively in
Table~\ref{table:T_isentrop}.  When
starting from a low temperature, we see almost no heating in
units of $t$ up to $U/t = 70$. The initially low entropy can be distributed over the remaining superfluid regions. %and the particle-hole excitations of the Mott phase.  
Measuring the temperature in units of $U$ leads to a cooling of the system
below the value of $U/t = 70$.

In contrast, for
interaction strengths larger than $U/t = 70$ almost the whole system is
occupied by a commensurate region and the incommensurate regions have a
negligible volume fraction. The energy cost to generate an
excitation in this situation corresponds in the bulk to the large interaction
energy and at the boundaries to the large potential energy cost resulting from
the steep trapping potential. Hence the entropy cannot be well accommodated in the system and the temperature in units of $t$
increases stronger than before when going to
$U/t=100$. In units of $U$ the temperature first drops before it increases again or saturates for large $U/t$
lattice potential. 

Starting with a low temperature ensures
that the temperature remains low enough for the presence of the
Mott-insulating regions. In contrast when starting from a hot temperature (the right-most example in Table~\ref{table:T_isentrop} is already in the normal state), there is heating in units of $t$ and only weak cooling in units of
$U$. The quantum degeneracy regime is never reached. Note that the description
by the one-band Hubbard model breaks down at such high temperatures.

%************************************************************************************************
%
%************************************************************************************************

\section{Interpretation of experimental results}
\label{sec:visibility}

One of the standard experimental observation techniques consists of suddenly
switching off the confinement and taking absorption images of the freely
expanding gas after a finite flight time. The hereby resulting interference pattern is a reflection of the initial momentum distribution,
\begin{equation}
n(\vec{k}) = \vert w(\vec{k}) \vert^2 \sum_{i,j}  \langle b_i^{\dagger} b_j \rangle e^{i \vec{k} \cdot(\vec{r}_i -  \vec{r}_j)},
\label{eq:mom}
\end{equation}
where the factor $w(\vec{k})$ is the Fourier transform of the Wannier function \cite{Jaksch98}. The use of the momentum profiles (and also of the visibilities) has been subject to debate, since in this quantity different effects as heating or the loss of coherence by stronger interactions have the same consequence. Furthermore,  it is very difficult to extract 
information on the superfluid-Mott insulator
transition point from these measurements in a trapping potential \cite{Wessel04, PhD, KollathZwerger2004}. 
This is due to the spatially coexisting regions of superfluid and
Mott-insulating character. A local measurement has to be implemented to obtain
detailed information about the system. A first step into this direction has
been taken by F\"olling {\it et al.} \cite{Foelling06} who measured the density
in thin slices. Experimental progress was reported for a local measurement of
the density in Ref.~\cite{Greinerxxx}, and another proposal was made
in Ref.~\cite{Kollath07}. Such measurement techniques should be preferred over the visibility, which is only well suited for identifying the Mott and superfluid phase far away from the transition point. 
In Figs.~\ref{fig:adiab_dens_s10} and~\ref{fig:adiab_compr_s10} we show the typical evolution of the density profiles (with Mott plateaus for strongly repulsive systems) and variance profiles along isentropic lines. 
 
\subsection{Dependence of visibility on temperature and trapping potential}

Before we compare our results to the experimentally extracted quantities we
would like to point out some features of the momentum distribution in a trapped
systems at finite temperature.
To do so we define the visibility $V$ 
\begin{equation}
V = \frac{n_{\rm max} - n_{\rm min}}{n_{\rm max} + n_{\rm min}}.
\end{equation} 
where $n_{\rm min}$ and $n_{\rm max}$ are the values of the largest and smallest value of $n(\vec{k})$.

Deep in the superfluid phase the darkest spot has almost $n_{\rm min} \approx
 0$, leading to a visibility close to unity, while in the Mott insulating
 phase the contrast between the brightest and the darkest spot is almost zero
 and one expects a very low visibility.

\begin{figure}
\centerline{\includegraphics[angle=270,width=\columnwidth ]{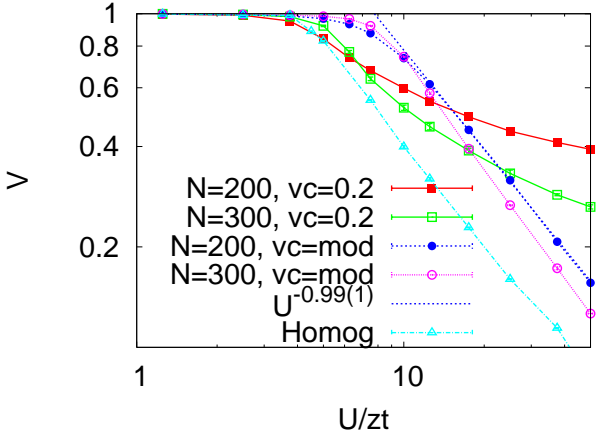}} 
\caption{ Visibility for a two-dimensional system $L=20 \times 20$ as a function of $U/zt$ at a high temperature of $\beta t = 0.4$ for different fillings ($N=200$ and $N=300$) and trapping potentials. $z$ is the coordination number. With $vc = 0.2$ we mean constant trapping $v_c/t = const$, while $vc = mod$ has parameters according to Table~\ref{table:pet_params}. The dashed line ('$U^{-0.99(1)}$') is a power-law fit with exponent $-1$ within error bars. Comparison with the homogeneous system at the same temperature and on a lattice of the same size is made ('Homog').  Its slope is equally -1. } \label{fig:visib_U}
\end{figure}

In Fig.~\ref{fig:visib_U} we show the visibilities for a trapped
two-dimensional Bose-Hubbard model at different densities. The calculation was done at a rather high
temperature $\beta t = 0.4$. Even at this temperature regions with integer
density and reduced compressibility exist. Precursors of the Mott-insulating
regions can be seen in the density distribution and its variance around $U/t \sim 25$. Looking
first at the case of a constant trapping potential, we see that for the chosen
number of particles a higher
density leads to a higher visibility at low  $U/t$. The visibility is lower at high  $U/t$ since the Mott region is larger. Taking the steepening of the trap into account
we see that at low values of $U/t$ the visibility is higher than the one obtained
with the constant trapping for the considered particle number. The reason is the increased number of particles
with density between one and two which form a superfluid edge between an $n=1$
and an $n=2$ Mott region. The $n=2$ Mott region is absent in the calculation with
the constant trapping potential.  For the two curves labeled 'vc=mod' taking the
change in the trapping potential into account, the visibility is well fitted by a
logarithmic curve for large values of $U /z t >~10$ in agreement to the finding in Ref. \cite{Gerbier07}. The reason
for this good agreement might be the suppression of the volume fraction of the superfluid regions, such
that only the Mott-insulating regions contribute to the decay of the visibility.

\begin{figure}
\centerline{\includegraphics[angle=270,width=\columnwidth ]{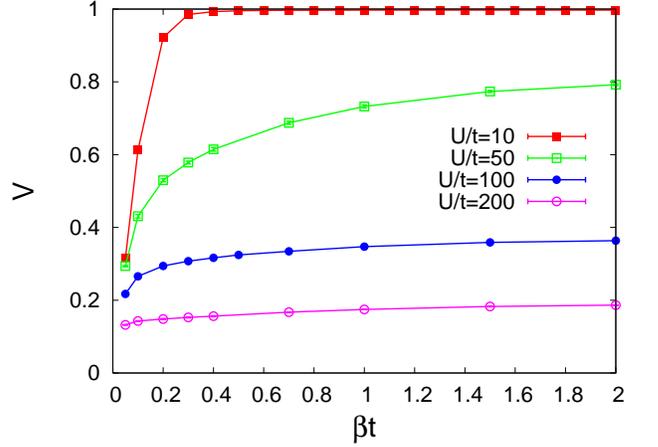}} 
\caption{ Visibility for a two-dimensional system $L=20 \times 20$ as a
  function of $\beta t$ at constant particle number $N=200$ for different values
  of $U/t$ and $v_c/t$ according to Table~\ref{table:pet_params}. } \label{fig:visib_beta}
\end{figure}

In Fig.~\ref{fig:visib_beta} we show the visibility as a function of the
inverse temperature for different strengths of the optical lattice
potential. We see a strong decrease in the visibility if the temperature becomes
of the order of $\beta t \approx 0.2$ when the quantum regime is left. A comparable decrease in the visibility at low temperature by changing the interaction
strength is only possible when it's very large, i.e. $U/t=200$ leads to a visibility of the order of $0.2$.
%% We see that the changes in the visibility are stronger when the quantum regime
%% is left rather than when the localization (i.e., $U/t$) is made stronger. For
%% interaction strength up to $U/t \approx 200$ we see that very low
%% values of the visibility are only found when the temperature is above $\beta
%% t\approx 0.2$
However, we should note that at high temperatures the single-band approximation of Eq.~(\ref{eq:bose_hubbard_hamiltonian}) loses its validity. \\
%Thus, for $\beta \le \left( (v_c)((L/2)^2+(L/2)^2) \right)^{-1}  \approx 0.025 $ the visibility is probably even lower than the values shown in Fig.~\ref{fig:visib_beta}.

\begin{figure}
\centerline{\includegraphics[angle=270,width=\columnwidth ]{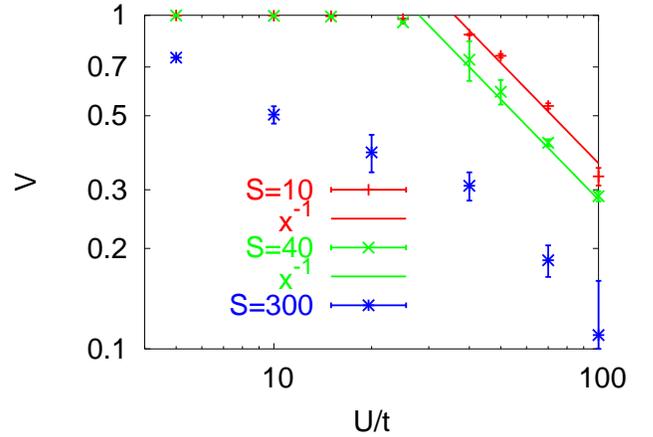}} 
\caption{Visibility along the entropic lines of Table~\ref{table:T_isentrop}. The visibility is computed by taking the brightest and darkest point of the momentum distribution of the Bose-Hubbard model only. } 
\label{fig:adiab_visib}
\end{figure}

 We further show the visibility along different isentropic lines in
 Fig.~\ref{fig:adiab_visib} for the same parameters as taken in
 Table~\ref{table:T_isentrop}. The curves with low entropy (low initial
 temperature) both have a slope of approximately $-1$, but the onset value
 $U/t$ is different. Even for the curve with high entropy, corresponding to an
 initial normal state, the data points for large value of $U/t$ seem to be
 consistent with this slope, but differ for a small value of $U/t$ (experiments show a constant visibility within error bars at low values of $U/t$). \\

\subsection{Comparison to experimental results}

We will now take the trap and the isentropic temperature change
into account, and compare with experiments. This is the hitherto most
realistic description done without numerical approximations, but for the 2D case we fail
to take the same number of atoms and system sizes as in the experiment. Further we
ignore time-of-flight collision effects.

\begin{figure}
\centerline{\includegraphics[angle=270,width=\columnwidth ]{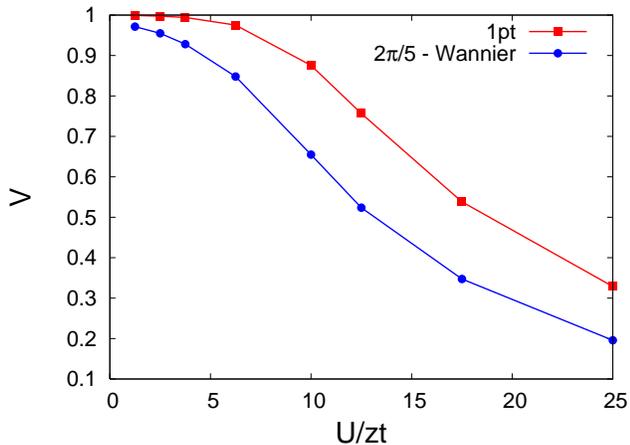}} 
\caption{Visibility $V$ as a function of $U/zt$ along an isentropic line
  $S=10$ with the same parameters as in Fig.~\ref{fig:entropy_pet} (see
  Table~\ref{table:pet_params}). We compare the ideal theoretical value
  obtained by taking just the maximum and minimum in the momentum profile
  ('1pt') %with visibility of the Bose-Hubbard model after averaging over a
	  %square of length $2\pi/5$ ('$2\pi/5$') and 
with the visibility after averaging over a square of length $2\pi/5$ taking
  the Wannier momentum profiles into account ('$2\pi/5 - {\rm Wannier}$'). The
  error bars are smaller than the point size. }
    \label{fig:visib_shell}
\end{figure}

We first compare our 2D calculations to the experimentally obtained
visibilities. Up to now we have evaluated the visibility using $n(\vec{k})$ at specific wave vectors $\vec{k}$.
In experiments instead an average over a square 
around the brightest and darkest spot is taken ( see  Ref.~\cite{Gerbier05PRL} for the exact experimental procedure). The size of the area
is a trade-off between signal and noise. The squares are chosen such that the
contribution of the envelope of the Wannier functions (cf. Eq. (\ref{eq:mom})) is
minimal, but not negligible. In Fig.~\ref{fig:visib_shell} we show the
different curves including the average and our previous definition
of the visibility.  For low values of
the lattice height the curves without the Wannier envelope have
values very close to one, whereas the curve considering the Wannier envelope
starts at a smaller value for the visibility. Fitting the slopes of the visibility with a power law gives roughly the same exponent for the theoretical and the experimental procedure, but the quality of the fit is better for the theoretical procedure.

Looking at three-dimensional experimental data, many groups find a visibility close to unity until $U/zt \sim 6.5$, where $z$ denotes the coordination number. For larger values of $U/t$ the decrease in visibility is well approximated by $V \sim (U/zt)^{\nu}$. The Z{\"u}rich group~\cite{Stoeferle04, Guenter06} finds $\nu = -1.36(5)$, while the Mainz group~\cite{Gerbier05} finds $\nu \sim -0.98(7)$. The reason for the discrepancy is not fully understood. Similarly, the momentum distribution data on a log-log plot of the two-dimensional system studied in Ref.~\cite{Spielman07} were consistent with a behavior of ($t/U$) and thus with a $\nu \approx -1$.

\begin{figure}
\centerline{\includegraphics[angle=270,width=\columnwidth ]{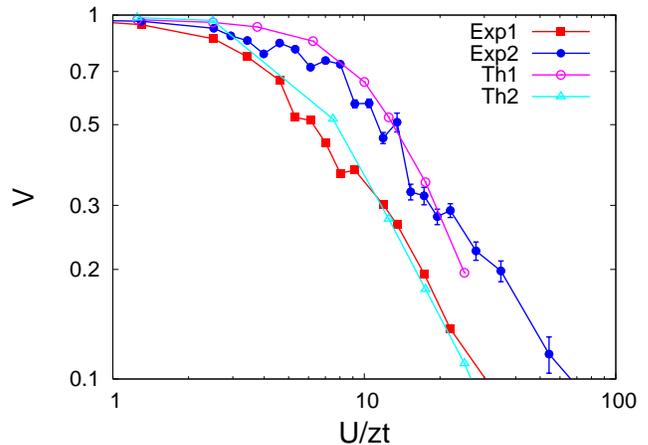}} 
\caption{Comparison between the 3d data of Ref.~\cite{Gerbier05PRL}
   (blue and red; the red data points have approximately 79,000 particles and the blue ones 224,000. This would correspond in the atomic limit to the presence of a small $n=2$ plateau for the lower particle number and to a $n=2$ plateau with almost half the number of particles for the higher number of particles\cite{private}.) and the visibility found in our  model
  (labeled 'Th1' and 'Th2'). The curve 'Th1' corresponds to the same parameters as in Fig.~\ref{fig:entropy_pet}
  along the adiabatic line $S=10$. The theoretical data is treated in the same way as
  explained in the text by taking the Wannier functions and averaging over a
  square of size $2 \pi/5$. The curve 'Th2' is a theoretical curve computed in the same way as 'Th1' and for the same system along an isentropic line $S = 5$ but for $N = 50$ particles such that the maximum occupation in the center of the trap does not exceed unity.  }
\label{fig:visib_bloch}
\end{figure}

In Fig.~\ref{fig:visib_bloch} we show a comparison between our theoretical results
 for a two-dimensional system and an example of the experimental data
 \cite{Gerbier05} taken for a three
 dimensional system. 
We rescale the interaction strength by the
 coordination number taking the mean-field effect between the different
 dimensions into account. We compare two theoretical curves with $N=50$ and $N=200$ particles. In the strong interaction limit this corresponds to a density profile which has an $n=1$ plateau for $N=50$ and a profile with an $n=1$ and an $n=2$ plateau for $N=200$. The initial temperatures at $U/t=5$  are $\beta t = 1.7$  and $\beta t=1.4$, respectively. Note that the particle numbers and temperature do not directly correspond to the experimental ones. For a direct comparison of an experimental to a theoretical curve, three dimensional simulation with realistic particle numbers would be needed. However, we find a good agreement in Fig.~\ref{fig:visib_bloch} of the theoretical
 visibility treated in the way described above with
 the experimental data even up to relatively strong optical lattices. This
 shows that the experimental visibility can be explained by just taking the
 isentropic change of the temperature into account. In particular, the system
 does not have to leave the quantum regime to reach the low values
 of the visibility.  The drop in the visibility can be explained by the
 formation of a broad Mott-insulating region (cf.~Fig.~\ref{fig:adiab_dens_s10} and
 \ref{fig:adiab_compr_s10}). 
 
 %Although our theoretical curve seems to have a steeper slope than the experimental curves for big values of $U/zt$, we believe that this is an artifact of our small system size since we overestimate the volume fraction affected by the big trapping confinement.\\

%%%%%%%%%%%%%%%%%%%%%%%%%%%%%%%%%%1D%%%%%%%%%%%%%%%%%

             In one-dimensional systems the procedure to get a quantity
  similar to the visibility has to be changed, since the interference
  pattern consists out of stripes. In Ref.~\cite{Stoeferle04} the superfluid to Mott-insulating
  transition in one-dimensional tubes was considered. The experimentalists extracted the
  coherent fraction of the atoms by taking the ratio of the content in the interference
  peaks and the background. We compared  (not shown) the experimental data with the
  theoretical calculations where we treated the results for the momentum
  distribution in the same way.  We found the same order of magnitude and the
  same qualitative trends ( e.g., the visibility for low lattice heights is
  between 0.5 and 0.6 in both cases). A detailed comparison however is hindered by the
  presence of many different parallel tubes in the experiments and by the
  large uncertainties stemming from the difficulty to fit the very sharp interference peaks.

\section{Summary}
\label{sec:conclusion}

An interpretation of experimental results on the superfluid to Mott insulator transition taking finite temperature effects into account has been given before by several authors, with differing conclusions. We have addressed the possibility of adiabatic temperature changes when ramping up the
lattice for a Bose-Hubbard model using unbiased and first-principle quantum
Monte Carlo simulations. We find that the expressions 'heating' and 'cooling' have to be
taken with care, since they strongly depend on the unit in which temperature
is measured. 

For the homogeneous case, we found some small heating in units of $t$  near
commensurability in the superfluid phase for high temperatures $\beta t <
0.5$. For low temperatures, there is some small heating for low values of $U/t <
(U/t)_c$, but the system cools slightly for larger values of $U/t$ when we are in
the proximity of commensurability. At very low densities, temperature remains
almost constant in the superfluid phase. When the density is commensurate,
the temperature shoots up dramatically in units of $t$ when the Mott gap opens in order to
keep the accessible number of levels constant.  This is in agreement
with the findings of previous studies \cite{ Reischl05, Schmidt06, Rey06}.

In a trapped system however the situation is different as already noted in
\cite{Rey06} for hard-core bosons. We find that in a one- and a two-dimensional system in a trapping
potential the main entropy contribution comes from incommensurate regions with
low filling $n\approx 0.5$. This is in good agreement with the finding for a three-dimensional system of the Amherst
group where they find that the non-commensurate edges between the Mott plateaus become
normal and accommodate the entropy~\cite{Capogrosso07}.
We found that the temperature increased in
units of $t$ when the size of the commensurate regions is broader than several lattice sites in the case of constant trapping. For
realistic trapping and densities around one or two, we found a temperature increase in units of
$t$, even in the weakly interacting regime.  However, in units of $U$ temperature
decreases or saturates, and its value lies deep in the quantum regime for the commensurate regions.  Only if the
incommensurate regions around filling $n\approx 0.5$ are almost totally suppressed the entropy has
to be taken by excitations in the commensurate regions or regions with higher filling, by which the
Mott-insulating regions might be destroyed. Our result is in agreement with the results of F. Gerbier who finds that current experiments easily reach the thermal insulator regime ($T < T^* \approx 0.2U$, where Mott-like features persist but superfluidity is absent), and possibly the quantum region~\cite{Gerbier07b}. In contrast to the predictions of Ref.~\cite{Ho07}, we find that for realistic paramters no runaway temperature occurs.  

Assuming adiabaticity in current experiments is
in agreement with theory remaining in the quantum regime. In
particular, the drop in the visibility of the interference pattern can be fully
explained within this framework and no additional temperature rise has to be
taken into account.

%This supports the view that in current experiments the quantum regime is
%reached. 
Future experiments using the spatially resolved measurements of the
density and higher order correlations will be able to confirm the creation of
Mott-insulating regions as the first evidence was reported in
\cite{FoellingBloch2006}. Unfortunately, we have seen that the integrated density fluctuations are not very sensitive to temperature changes when one is deep in the quantum regime.

\section{Outlook and Conclusions}
Our principal assumption that the loading of the lattice is approximately adiabatic needs to be verified considering the dynamical process at finite temperature. 
It will be equally important to extend our investigations to different and larger systems. A major goal will be to treat the full three-dimensional Bose-Hubbard system with a realistic number of particles. Systems with different types of particles will have to be addressed as well, since the development of a new energy scale that does not scale with $U$ will have a negative influence on the possibility of reaching the ground state adiabatically. The visibility results of theory~\cite{Pollet06} at low temperature and experiment~\cite{Guenter06, Ospelkaus06} are in disagreement for Bose-Fermi mixtures. It was believed that temperature is such that these systems are not in their ground state~\cite{Pollet06}, and a study by Cramer {\it et al.}~\cite{Cramer07} hints at heating effects for a weak inter-species coupling. %Spin systems (with a Neel ground state) might also be studied in optical lattices, but Ref.~\cite{Koetsier07} puts severe limitations on the possibility of reachin
 g the ground state. 
Anti-ferromagnetism and entropy were previously addressed for a homogeneous Fermi-Fermi system in Ref~\cite{Werner05, Dare07,Koetsier07}.

This work was motivated by the strongly different opinions that existed about temperature effects in the Bose-Hubbard model, ranging from a constant temperature to a runaway temperature. We have addressed all aspects of this discussion in homogeneous and trapped, one and two-dimensional Bose-Hubbard systems and the results of our work are uni-vocal: compatibility of experimental with numerical  visibility curves (and density profiles) supports that the experimental initial temperature of the BEC is deeply in the superfluid phase, and that the quantum regime is not left even when adiabatically ramping up the optical lattice, even in the presence of a considerable Mott domain. The temperature scales almost linearly with $U$ when the  Mott domain is considerable in size, but temperature remains a very low fraction of the value of $U$. We have copied the experimental procedure of measuring the visibility in our simulations and found good agreement. It turns out that  the Wannier function and the averaging over a small region of the interference pattern produce an almost constant shift of the visibility as a function of $U/t$. For deep lattices, the trapping potential becomes effectively steeper, and different Mott plateaus are forced to form. The main contributions to the visibility come from particle-hole excitations giving the visibility a slope of $-1$ as a function of $U/t$. The superfluid volume fraction is too low, but it still absorbs most of the entropy, as we could infer from the local density approximation. Our results strongly support that experimentalists have observed the superfluid and Mott phase and their crossover in a trapped system.

We wish to thank I. Bloch, T. Esslinger, A. Georges, T. L. Ho, S. Huber, H. Moritz, J. V. Porto, and S. Wessel for fruitful discussions.
We thank H. Moritz, F. Gerbier, I. Bloch and I. Spielman for providing us the experimental data.
We wish to thank the Aspen Center for Physics and the Institute Henri Poincare-Centre Emile Borel for hospitality and support.
We thank the Swiss National Science Foundation, MaNEP Switzerland, the CNRS, the RTRA network 'Triangle de la Physique', the DARPA OLE program and FWO Vlaanderen for financial support.
Part of the simulations ran on the Hreidar cluster at ETH Zurich.

\end{document}